\documentclass[12pt]{article}
\usepackage{a4wide}
\usepackage{epsfig}
\newcommand{\oone}{\hbox{$1\kern-2.5pt\hbox{\rm l}$}}
\newcommand{\ssigma}{\hbox{$\kern2.5pt\vrule height4pt\kern-2.5pt\sigma$}}

\newcommand{\GeV}{{\rm\,GeV}}
\newcommand\pfrac[2]{\left(\frac{#1}{#2}\right)}
\newcommand{\Li}{{\rm Li}}
\newcommand{\imag}{\mathop{\rm Im}\nolimits}
\newcommand{\real}{\mathop{\rm Re}\nolimits}

\begin{document}

\thispagestyle{empty} 
\begin{flushright}
MZ-TH/09-18\\
0905.4465 [hep-ph]\\
\end{flushright}
\vspace{0.5cm}

\begin{center}
{\Large\bf $O(\alpha_s)$ corrections to the polar angle\\[.2cm]
  dependence of the longitudinal spin--spin\\[.3cm]
  correlation asymmetry in $e^+e^-\to q\bar q$}\\[1.3cm]
{\large S.~Groote$^{1,2}$, J.G.~K\"orner$^1$ and J.A.~Leyva$^1$}\\[1cm]
$^1$ Institut f\"ur Physik, Johannes-Gutenberg-Universit\"at,\\[.2cm]
Staudinger Weg 7, 55099 Mainz, Germany\\[7pt]
$^2$ Loodus- ja Tehnoloogiateaduskond, F\"u\"usika Instituut,\\[.2cm]
  Tartu \"Ulikool, T\"ahe 4, 51010 Tartu, Estonia
\end{center}

\vspace{1cm}
\begin{abstract}\noindent
We provide analytical results for the $O(\alpha_s)$ corrections to the polar
angle dependence of the longitudinal spin--spin correlation asymmetry in 
$e^+e^-\to q\bar q$. For top quark pair production the $O(\alpha_s)$
corrections to the longitudinal spin--spin asymmetry are strongly polar angle
dependent and can amount up to $\simeq 4\%$ in the $q^2$-range from above
$t\bar t$ threshold up to $\sqrt{q^2}=1000\GeV$. The $O(\alpha_s)$ radiative 
corrections to the correlation asymmetry are below $\approx 1\%$ in the 
forward direction where the cross section is largest. In the
$e^+e^-\to b\bar b$ case the $O(\alpha_s)$ corrections reduce the asymmetry
value from its $m_b=0$ value of $-100\%$ to approximately $-96\%$ for
$q^2$-values around the $Z$ peak and are practically independent of the value
of the polar angle $\theta$. This reduction can be traced to finite anomalous
contributions from residual mass effects which survive the $m_b\to 0$ limit.
We discuss the role of the anomalous contributions and the pattern of how they
contribute to spin-flip and non-flip terms.
\end{abstract}

\newpage


\section{Introduction}


Since polarized top and antitop quarks decay weakly before hadronizing they
retain their original polarization from the production process when they
decay. There are thus significant correlations between the decay products of
jointly produced top and antitop quarks. The relevant spin information is
encoded in the spin--spin density matrix of the $(t\bar t)$-system. For
hadronically produced $(t\bar t)$-pairs such correlations have been thoroughly
discussed in the review article~\cite{Bernreuther:2008ju}. In
$e^+e^-\to t\bar t$ interactions the spin--spin density matrix of the
$(t\bar t)$-system~\cite{ps96,Akatsu:1997tq,OlsenStav,TBP2,GKL2,%
Brandenburg:1998xw} and the ensuing correlations between the decay
products~\cite{ps96,Akatsu:1997tq} have been studied in a number of papers.

Of interest is also the role of quark mass effects in the production of quarks
and gluons in $e^+e^-$-annihilations. Jet definition schemes, event shape
variables, heavy flavour momentum correlations~\cite{BBU,Rodrigo,Nason} and the
polarization of the gluon~\cite{GKL1} are affected by the presence of quark
masses for charm and bottom quarks even when they are produced at the scale of
the $Z^0$ mass. A careful investigation of quark mass effects in
$e^+e^-$-annihilations may even lead to an alternative determination of the
quark mass values~\cite{BBU,Rodrigo,Nason,Grunberg}. There is an obvious
interest in quark mass effects for $(t\bar t)$-production where quark mass
effects cannot be neglected in the envisaged range of energies to be covered
by the proposed International Linear Collider (ILC). At the low mass end quark
mass effects are important in the $m_q\to 0$ calculation of radiative
corrections to quark polarization variables because residual mass effects
change the naive non-flip pattern of the $m_q=0$ polarization
results~\cite{KPT,GKT2}. In QED the same $O(\alpha/\pi)$ residual mass effects
have been widely discussed in polarized lepton
production~\cite{LeeNau,Smi,Was:1986dv} and polarization effects in lepton
decays~\cite{Falk,Fischer:2002hn,trentadue}.

In this report we provide analytical results for the $O(\alpha_s)$ radiative 
corrections to the longitudinal spin--spin correlation asymmetry (called
longitudinal spin--spin asymmetry for short) and its polar angle dependence
for massive quark pairs produced in $e^+e^-$-annihilations. This goes beyond
the calculation of Ref.~\cite{TBP2,GKL2} where the polar angle dependence was
averaged over. In contrast to the analysis \cite{Brandenburg:1998xw}, which was
numerical, our results are presented in closed analytic form, which allows
one to explicitly study the high energy (or $m_{q} \to 0$) limit of the
relevant spin--spin density matrix elements which differ from the naive
$m_q=0$ values due to the residual quark mass effects.

The longitudinal polarization of massive quark pairs produced in
$e^+e^-$ interactions affects the shape of the energy spectrum of their
secondary decay leptons. For example, the longitudinal spin--spin correlation
effects in pair produced top quarks and antitop quarks will lead to
correlation effects of the energy spectra of their secondary decay leptons and
antileptons. We discuss in detail the $m_q\to 0$ limit of the relevant
polarized structure functions and the role of the $O(\alpha_s)$ residual mass
effects which are mostly relevant for charm and bottom quarks for energies at
and above the $Z$ and for top quarks at the highest energy
$\sqrt{q^2}=1000\GeV$. We delineate how residual mass effects contribute to
the various spin-flip and non-flip terms in the $m_q\to 0$ limit for each of
the three spin--spin correlation functions that describe the polar angle
dependence of the longitudinal spin--spin correlations.


\section{Joint quark-antiquark density matrix\\ and polar angle distribution}


Let us begin by defining the differential joint quark-antiquark density matrix
$d\ssigma^\alpha=d\sigma^\alpha_{\lambda_1\lambda_2;\lambda'_1\lambda'_2}$
where $\lambda_1$ ($\lambda'_1$) and $\lambda_2$ ($\lambda'_2$) denote the
helicities of the quark and antiquark, respectively. The label $\alpha$
specifies the polarization of the initial $\gamma^*$ and $Z$, or interference
contributions thereof, which determines the polar angle dependence of the
cross section and the longitudinal spin--spin correlations. In our notation the
three polarization components are called $\alpha=U$ (unpolarized transverse),
$L$ (longitudinal) and $F$ (forward--backward).

In this paper we concentrate on the longitudinal polarization of the quark and
antiquark, and in particular, on their longitudinal spin--spin correlations.
Thus we specify to the diagonal case $\lambda_1=\lambda'_1$ and
$\lambda_2=\lambda'_2$. The diagonal part of the differential joint density
matrix can be represented in terms of its components along the products of the
unit matrix $\oone$ and the $z$-components of the Pauli matrix $\ssigma_3$
($\ssigma_3=\hat p_1\vec{\ssigma}$ for the quark and
$\ssigma_3=\hat p_2\vec{\ssigma}$ for the antiquark,
$\hat p_i=\vec p_i/|\vec p_i|$). One has
\begin{equation}\label{spinspin1}
d\ssigma_\alpha=\frac14\left(\sigma_\alpha\oone\otimes\oone
  +\sigma_\alpha^{(\ell_1)}\ssigma_3\otimes\oone
  +\sigma_\alpha^{(\ell_2)}\oone\otimes\ssigma_3
  +\sigma_\alpha^{(\ell_1\ell_2)}\ssigma_3\otimes\ssigma_3\right)
\end{equation}
where the Pauli matrices to the left and right of the $\otimes$ symbol are
associated with the quark and the antiquark, respectively.

In order to get a feeling about the physical significance of the single and
double spin functions $\sigma_\alpha^{(\ell_{1,2})}$ and
$\sigma_\alpha^{(\ell_1\ell_2)}$ consider the case where the polarization of
the quarks and antiquarks is purely longitudinal in the helicity system. This
would be the case in the high energy limit since the transverse spin
components of the top and antitop quark go to zero as $m_q/\sqrt{q^2}$. The
differential distribution with respect to the polar angles $\cos\theta_{1,2}$
specifying the angles between the (quark, antiquark) and given decay products
thereof reads
\begin{eqnarray}
\frac{d\sigma_{\alpha}}{d\cos\theta_1d\cos\theta_2}&=&\frac14
\bigg(\sigma_{\alpha}+\sigma_\alpha^{(\ell_1)}\alpha_1\cos\theta_1
+\sigma_\alpha^{(\ell_2)}\alpha_2\cos\theta_2\nonumber\\&&\qquad
+\sigma_\alpha^{(\ell_1\ell_2)}\alpha_1\alpha_2\cos\theta_1\cos\theta_2\bigg)
\end{eqnarray}
where the asymmetry parameters $\alpha_{1,2}$ specify the analyzing power
of the particular final state in the respective decay channels. Consider, for
example, $(t\bar{t})$-production followed by the semileptonic decays
$t\to b+\ell^++\nu_\ell$ and $\bar{t}\to\bar{b}+\ell^-+\bar{\nu}_{\ell}$.
Each of the decay products of the top and antitop can serve as a polarization
analyzer of the spin of the top and antitop quarks. From the experimental
point of view the charged leptons $\ell^\pm$ are ideally suited for this
purpose because they are easy to detect and because their analyzing power is
maximal with $\alpha_{1,2}=\pm 1$ (see e.g.~\cite{Groote:2006kq}).

An alternative but equivalent representation of the longitudinal spin
contributions in Eq.~(\ref{spinspin1}) can be written down for the double
density matrix elements $\sigma_{\alpha}(s_1^\ell,s_2^\ell)$ in terms of the
longitudinal spin components $s_1^\ell=2\lambda_1$ and $s_2^\ell=2\lambda_2$
with $s_1^\ell,s_2^\ell=\pm 1$ (or
$s_1^\ell,s_2^\ell\in\{\uparrow,\downarrow\}$). One has
\begin{equation}\label{eqnkd1}
\sigma_{\alpha}(s_1^\ell,s_2^\ell)=\frac14\left(\sigma_\alpha
+\sigma^{(\ell_1)}_\alpha s_1^\ell+\sigma^{(\ell_2)}_\alpha s_2^\ell
+\sigma^{(\ell_1\ell_2)}_\alpha s_1^\ell s_2^\ell\right).
\end{equation}
Eq.~(\ref{eqnkd1}) is easily inverted. The result is
\begin{eqnarray}
\label{uparrow1}
\sigma_\alpha&=&\sigma_\alpha(\uparrow\uparrow)
  +\sigma_\alpha(\uparrow\downarrow)+\sigma_\alpha(\downarrow\uparrow)
  +\sigma_\alpha(\downarrow\downarrow),\nonumber\\
\sigma_\alpha^{(\ell_1)}&=&\sigma_\alpha(\uparrow\uparrow)
  +\sigma_\alpha(\uparrow\downarrow)-\sigma_\alpha(\downarrow\uparrow)
  -\sigma_\alpha(\downarrow\downarrow),\nonumber\\
\sigma_\alpha^{(\ell_2)}&=&\sigma_\alpha(\uparrow\uparrow)
  -\sigma_\alpha(\uparrow\downarrow)+\sigma_\alpha(\downarrow\uparrow)
  -\sigma_\alpha(\downarrow\downarrow),\nonumber\\
\sigma_\alpha^{(\ell_1\ell_2)}&=&\sigma_\alpha(\uparrow\uparrow)
  -\sigma_\alpha(\uparrow\downarrow)-\sigma_\alpha(\downarrow\uparrow)
  +\sigma_\alpha(\downarrow\downarrow).
\end{eqnarray}
In the following we shall refer to the $(\uparrow\downarrow)$ and
$(\downarrow\uparrow)$ spin configurations as aligned even if the quark and
antiquark are not parallel.

$O(\alpha_s)$ radiative corrections to the unpolarized rate components
$\sigma_\alpha$ have been discussed before (see e.g.~\cite{KPT,GKT2,TBP1})
including beam polarization effects~\cite{GKT2} and beam-event correlation
effects~\cite{GKT2,TBP1}. The $O(\alpha_s)$ radiative corrections to the
longitudinal spin component $\sigma_{U+L}^{(\ell_1)}$ (and thereby
$\sigma_{U+L}^{(\ell_2)}$) have been calculated in
Refs.~\cite{KPT,Kodaira:1998gt}. Beam polarization and $O(\alpha_s)$
beam-event correlation effects for $\sigma_{\alpha}^{(\ell_1)}$
($\alpha=U,L,F$) were calculated in~\cite{GKT2,GKT1,Ravindran:2000rz}. As
concerns the spin--spin correlation asymmetry $\sigma_\alpha^{(\ell_1\ell_2)}$
the $O(\alpha_s)$ tree-graph contributions have been written down in
Ref.~\cite{OlsenStav}. The complete $O(\alpha_s)$ radiative corrections to the
fully integrated spin--spin correlation component
$\sigma_{U+L}^{(\ell_1\ell_2)}$ were calculated in Refs.~\cite{TBP2,GKL2}
where beam-event correlation effects were averaged over. These results were
confirmed in a numerical calculation by
Brandenburg {\it et al.}~\cite{Brandenburg:1998xw}. As mentioned in the
introduction to this paper we are aiming to determine the polar angle
dependence of the longitudinal spin--spin correlation asymmetry, i.e.\ we are
interested in the polar angle structure induced by the rate functions
$\sigma_\alpha$ and the longitudinal spin--spin functions
$\sigma_\alpha^{(\ell_1\ell_2)}$ ($\alpha=U,L,F$).

As before we write the electroweak cross section and the spin--spin
correlation components in modular form in terms of three building
blocks~\cite{GKT2}, namely the lepton tensor (which encapsules the beam
polarization parameters and the angular dependences), the hadron tensor (which
contains the hadron dynamics) and an electroweak coupling matrix $g_{ij}(q^2)$
which connects the two. Since the electroweak model dependence and the polar
angle dependence of $\sigma_\alpha$ and $d\sigma_\alpha^{(\ell_1\ell_2)}$ are
the same, we introduce a compact notation $\sigma_\alpha^{\{\ell_1\ell_2\}}$
where $\sigma_\alpha^{\{\ell_1\ell_2\}}$ stand for either $\sigma_\alpha$ or
$\sigma_\alpha^{(\ell_1\ell_2)}$. Thus we write
\begin{eqnarray}\label{sigma1}
\frac{d\sigma^{\{\ell_1\ell_2\}}}{d\cos\theta}
  &=&\frac38(1+\cos^2\theta)\left(g_{11}\sigma_U^{1\,\{\ell_1\ell_2\}}
  +g_{12}\sigma_U^{2\,\{\ell_1\ell_2\}}\right)\nonumber\\&&
  +\frac34\sin^2\theta\left(g_{11}\sigma_L^{1\,\{\ell_1\ell_2\}}
  +g_{12}\sigma_L^{2\,\{\ell_1\ell_2\}}\right)\nonumber\\&&
  +\frac34\cos\theta\left(g_{43}\sigma_F^{3\,\{\ell_1\ell_2\}}
  +g_{44}\sigma_F^{4\,\{\ell_1\ell_2\}}\right).
\end{eqnarray}
The index $j=1,2,3,4$ \,in $\sigma_\alpha^{j\,\{\ell_1\ell_2\}}$ runs over the
four linear combinations of bilinear products of vector and axial vector
currents defined by
\begin{eqnarray}\label{H1234}
\sigma_\alpha^1&=&\frac12(\sigma_\alpha^{VV}+\sigma_\alpha^{AA}),\qquad
\sigma_\alpha^2\ =\ \frac12(\sigma_\alpha^{VV}-\sigma_\alpha^{AA}),\nonumber\\
\sigma_\alpha^3&=&\frac{i}2(\sigma_\alpha^{VA}-\sigma_\alpha^{AV}),\qquad
\sigma_\alpha^4\ =\ \frac12(\sigma_\alpha^{VA}+\sigma_\alpha^{AV}).
\end{eqnarray}
$\theta$ is the polar angle between the electron beam direction and the top
quark direction.

The matrix $g_{ij}$ ($i,j=1,2,3,4$) specifies the electroweak model
dependence of the $e^+e^-$ cross section. For the present discussion we only
need the components $g_{11}$, $g_{12}$, $g_{43}$ and $g_{44}$. They are given
by
\begin{eqnarray}\label{gdiag}
g_{11}&=&Q_f^2-2Q_fv_ev_f\real\chi_Z
  +(v_e^2+a_e^2)(v_f^2+a_f^2)|\chi_Z|^2,\nonumber\\
g_{12}&=&Q_f^2-2Q_fv_ev_f\real\chi_Z
  +(v_e^2+a_e^2)(v_f^2-a_f^2)|\chi_Z|^2,\nonumber\\
g_{43}&=&2Q_fa_ea_f\imag\chi_Z,\nonumber\\
g_{44}&=&-2Q_fa_ea_f\real\chi_Z+4v_ea_ev_fa_f|\chi_Z|^2
\end{eqnarray}
where for the Standard Model $\chi_Z(q^2)=gM_Z^2q^2/(q^2-M_Z^2+iM_Z\Gamma_Z)$
with $M_Z$ and $\Gamma_Z$ the mass and width of the $Z^0$ and
$g=G_F(8\sqrt2\pi\alpha)^{-1}\approx4.49\times 10^{-5}\GeV^{-2}$. $Q_f$ are the
charges of the final state quarks to which the electroweak currents directly
couple, $v_e$ and $a_e$, $v_f$ and $a_f$ are the electroweak vector and axial
vector coupling constants. For example, in the Weinberg--Salam model one has
$v_e=-1+4\sin^2\theta_W$, $a_e=-1$ for leptons, $v_f=1-\frac83\sin^2\theta_W$,
$a_f=1$ for up-type quarks ($Q_f=\frac23$), and $v_f=-1+\frac43\sin^2\theta_W$,
$a_f=-1$ for down-type quarks ($Q_f=-\frac13$). At low energies, where only
QED interactions survive, the only remaining components of the electroweak
coupling matrix $g_{ij}$ are $g_{11}=Q_f^2$ and $g_{12}=Q_f^2$. In this
paper we use Standard Model couplings with $\sin^2\theta_W=0.231$ and
$M_Z=91.188\GeV$, $\Gamma_Z=2.487\GeV$~\cite{PDG08}.

The hadronic building block is entirely determined by the hadron dynamics,
i.e.\ by the current-induced production of a quark-antiquark pair which, in
the $O(\alpha_s)$ case, is followed by gluon emission. At $O(\alpha_s)$
one also has to add the one-loop contribution. We shall work in terms of
unpolarized and polarized hadron tensor components $H_\alpha^j$ and
$H_\alpha^{j\,(\ell_1\ell_2)}$ where the spin decomposition is defined in
complete analogy to Eq.~(\ref{eqnkd1}). The notation closely follows the
notation used in Ref.~\cite{GKT2}.

In the two-body case $e^+e^-\to q(p_1)\bar q(p_2)$ (Born and one-loop
contribution) the unpolarized rate components $\sigma_\alpha^j$ and the
longitudinal spin--spin components $\sigma_\alpha^{j\,(\ell_1\ell_2)}$ are
related to the corresponding two-body hadronic tensor components
$H_\alpha^{j\,\{\ell_1\ell_2\}}$ via
\begin{equation}\label{crosssection2}
\sigma_\alpha^{j\,\{\ell_1\ell_2\}}=\frac{\pi\alpha^2v}{3q^4}
  H_\alpha^{j\,\{\ell_1\ell_2\}}
\end{equation}
where $v^2=1-4m_q^2/q^2:=1-\xi$. The helicity structure functions
$H_\alpha^{j\,\{\ell_1\ell_2\}}$ ($\alpha=U,L,F$) are obtained from the hadron
tensor components $H_{\mu\nu}^{j\,\{\ell_1\ell_2\}}$ via covariant projection
\begin{eqnarray}\label{ULFproj}
H_U^{j\,\{\ell_1\ell_2\}}&=&\left(-\hat g_{\mu\nu}
  -\frac{\hat p_1^\mu\hat p_1^\nu}{p_{1z}^2}\right)
  H_{\mu\nu}^{j\,\{\ell_1\ell_2\}},\qquad
H_L^{j\,\{\ell_1\ell_2\}}\ =\ \frac{\hat p_1^\mu\hat p_1^\nu}{p_{1z}^2}
  H_{\mu\nu}^{\{\ell_1\ell_2\}},\nonumber\\
H_F^{j\,\{\ell_1\ell_2\}}&=&i\varepsilon^{\mu\nu\alpha\beta}
  \frac{\hat p_{1\alpha}q_\beta}{p_{1z}\sqrt{q^2}}
  H_{\mu\nu}^{j\,\{\ell_1\ell_2\}}
\end{eqnarray}
where $\hat g_{\mu\nu}=g_{\mu\nu}-q_\mu q_\nu/q^2$ and
$\hat p_{1\mu}=p_{1\mu}-(p_1\cdot q)q_\mu/q^2$ are the four-transverse metric
tensor and the four-transverse quark momentum, respectively. The two-body
hadron tensor $H_{\mu\nu}^{\{\ell_1\ell_2\}}$, finally, is defined by the
product of matrix elements
\begin{equation}\label{matrix2}
H^{j}_{\mu\nu}(p_1,p_2,s_1^l,s_2^l)
=\langle q(s_1^l)\bar q(s_2^l)|J^{V,A}_\mu|0\rangle
\langle 0|J_\nu^{\dagger \,V,A}|q(s_1^l)\bar q(s_2^l)\rangle,
\end{equation}
where, again, the index $j=1,2,3,4$ specifies the current composition in terms
of the parity-even (for $j=1,2$) and parity-odd ($j=3,4$) products of the
vector and the axial vector currents as defined in Eq.~(\ref{H1234}).

In the three-body case $e^+e^-\to q(p_1)\bar q(p_2)g(p_3)$ the hadron tensor
$H_{\mu\nu}^j$ is obtained from the square of the current matrix elements
according to
\begin{equation}\label{matrix3}
H_{\mu\nu}^j(p_1,p_2,p_3,s_1^l,s_2^l)=\sum_{\rm gluon\ spin}\langle q(s_1^l)
\bar q(s_2^l)g|J^{V,A}_{\mu}|0\rangle \langle 0|J_\nu^{\,\dagger\,V,A}
|q(s_1^l)\bar q(s_2^l)g\rangle.
\end{equation}
Note that the three-body hadron tensor in Eq.~(\ref{matrix3}) has a mass
dimension differing from that of the two-body hadron tensor in
Eq.~(\ref{matrix2}). The projection of the three-body hadron tensor onto the
six helicity structure functions $H_\alpha^{\{\ell_1\ell_2\}}$ ($\alpha=U,L,F$)
is done in identical fashion to the two-body case. What is finally needed is
the relation between the differential rate components and the three-body
helicity structure functions. The relation is given by
\begin{equation}\label{2to3}
\frac{d\sigma_\alpha^{j\,\{\ell_1\ell_2\}}}{dy\,dz}=\frac{\pi\alpha^2v}{3q^4}
  \left\{\frac{q^2}{16\pi^2v}H_\alpha^{j\,\{\ell_1\ell_2\}}(y,z)\right\}
\end{equation}
where we have explicitly referred to the phase space dependence of the
three-body hadron tensor in order to set it aside from the two-body hadron
tensor in Eq.~(\ref{matrix2}). As kinematic variables we use the two
energy-type variables $y=1-2p_1q/q^2$ and $z=1-2p_2q/q^2$.

We mention that the above formalism allows for a straightforward incorporation
of transverse and longitudinal beam polarization effects~\cite{GKT2}. For
example, if one starts with longitudinally polarized beams, one has to effect
the replacement
\begin{eqnarray}\label{ewcoupling2}
g_{1i}&\rightarrow&(1-h^-h^+)g_{1i}+(h^--h^+)g_{4i}\qquad(i=1,2)\nonumber\\
g_{4j}&\rightarrow&(1-h^-h^+)g_{4j}+(h^--h^+)g_{1j}\qquad(j=3,4)
\end{eqnarray}
where $h^-$ and $h^+$ ($-1\le h^\pm\le+1$) denote the longitudinal
polarization of the electron and the positron beam, respectively. Clearly
there is no interaction between the beams when $h^+=h^-=\pm1$. The additional
electroweak components $g_{41}$, $g_{42}$, $g_{13}$, and $g_{14}$ needed in
Eqs.~(\ref{ewcoupling2}) are given by
\begin{eqnarray}\label{gnondiag}
g_{41}&=&2Q_fa_ev_f\real\chi_Z-2v_ea_e(v_f^2+a_f^2)|\chi_Z|^2,\nonumber\\
g_{42}&=&2Q_fa_ev_f\real\chi_Z-2v_ea_e(v_f^2-a_f^2)|\chi_Z|^2,\nonumber\\
g_{13}&=&-2Q_fv_ea_f\imag\chi_Z,\nonumber\\
g_{14}&=&2Q_fv_ea_f\real\chi_Z-2(v_e^2+a_e^2)v_fa_f|\chi_Z|^2.
\end{eqnarray}
The incorporation of transverse beam polarization is described in
Ref.~\cite{GKT2}.


\section{Born-term contributions}


Let us begin by listing the non-vanishing Born-term contributions to the
unpolarized and the polarized spin--spin two-body hadron tensor components.
They can be obtained from the two-body hadron tensors in covariant fashion by
using the projections (\ref{ULFproj}) and the two-body representations of the
spin vectors (\ref{spin1}) and (\ref{spin2}). An alternative and rather
convenient approach is to use helicity amplitudes
$h^{V,A}_{\lambda;\lambda_q\lambda_{\bar q}}$ in the c.m.\ frame where
$\lambda$ denotes the rest frame $m$-quantum number of the $(\gamma^*,Z)$ (the
quark momentum defines the $z$-direction). The helicity amplitudes are given by
\begin{eqnarray}\label{helamp}
h^V_{0;\pm\frac12\pm\frac12}&=&-\sqrt{1-v^2}\,\sqrt{q^2},\qquad
h^A_{0;\pm\frac12\,\pm\frac12}= 0,\nonumber\\
h^V_{\pm 1;\pm\frac12\,\mp\frac12}&=& -\sqrt2\sqrt{q^2},\qquad\qquad
h^A_{\pm 1;\pm\frac12\mp\frac12}=\mp\sqrt2v\sqrt{q^2}.
\end{eqnarray}
One obtains
\begin{eqnarray}\label{uborn}
H_U^1({\it Born})&=&2N_cq^2(1+v^2),\nonumber\\
H_U^2({\it Born})&=&2N_cq^2(1-v^2),\nonumber\\
H_L^1({\it Born})&=&H_L^2({\it Born})\ =\ N_cq^2(1-v^2),\nonumber\\
H_F^4({\it Born})&=&4N_cq^2v
\end{eqnarray}
and
\begin{eqnarray}\label{pborn}
H_U^{1\,(\ell_1\ell_2)}({\it Born})&=&-2N_cq^2(1+v^2),\nonumber\\
H_U^{2\,(\ell_1\ell_2)}({\it Born})&=&-2N_cq^2(1-v^2),\nonumber\\
H_L^{1\,(\ell_1\ell_2)}({\it Born})&=&H_L^{2\,(\ell_1\ell_2)}({\it Born})
  \ =\ N_cq^2(1-v^2),\nonumber\\
H_F^{4\,(\ell_1\ell_2)}({\it Born})&=&-4N_cq^2v.
\end{eqnarray}

For completeness we also list the single spin Born term contributions.
They read
\begin{eqnarray}
H_U^{3\,(\ell_{1,2})}&=&0, \nonumber\\
H_U^{4\,(\ell_{1,2})}&=&\pm4N_Cq^2v,\nonumber\\
H_L^{3\,(\ell_{1,2})}&=&H_L^{4\,\ell_{1,2}}=0,\nonumber\\
H_F^{1\,(\ell_{1,2})}&=&\pm2N_Cq^2(1+v^2),\nonumber\\
H_F^{2\,(\ell_{1,2})}&=&\pm2N_Cq^2(1-v^2),
\end{eqnarray}
where the upper and lower signs stand for the quark and antiquark case,
respectively.\footnote {The signs of $H_F^{1,2\,\ell_2}$ differ from those
in Ref.~\cite{Groote:2008ux} because, contrary to~\cite{Groote:2008ux}, in
the present paper we are always referring to the same polar angle in
Eq.~(\ref{sigma1}) for both the quark and antiquark case.} Note that one has
the Born-term relations
\begin{eqnarray}\label{born1}
H_U^{1,2}({\it Born})&=&-H_U^{1,2\,(\ell_1\ell_2)}({\it Born}),\nonumber\\
H_F^4({\it Born})&=&-H_F^{4\,(\ell_1\ell_2)}({\it Born}),\nonumber\\
H_L^{1,2}({\it Born})&=&H_L^{1,2\,(\ell_1\ell_2)}({\it Born}).
\end{eqnarray}
These relations can be seen to arise from angular momentum conservation at the
Born-term level where the produced quark and antiquark are in a back-to-back
configuration. The easiest way to see this is by noting that at the Born level
one has $H_{U,F}(\uparrow\uparrow)=H_{U,F}(\downarrow\downarrow)=0$ and
$H_L(\uparrow\downarrow)=H_L(\downarrow\uparrow)=0$ from angular momentum
conservation. The above Born-term relations then follow from inserting the
above zero entries into Eqs.~(\ref{uparrow1}) (with $\sigma_\alpha$ replaced
by $H_\alpha$). It is quite clear that the relations Eq.~(\ref{born1}) no
longer hold true at $O(\alpha_s)$ in general since the quark and antiquark
are no longer back-to-back due to additional gluon emission.

Let us now introduce the normalized $\cos\theta$-dependent longitudinal
spin--spin asymmetry $P^{\ell\ell}(\cos\theta)$ which is
defined by
\begin{equation}\label{corrdef}
P^{\ell\ell}(\cos\theta)=\frac{\frac38(1+\cos^2\theta)\sigma_U^{(\ell_1\ell_2)}
  +\frac34\sin^2\theta\ \sigma_L^{(\ell_1\ell_2)}
  +\frac34\cos\theta\ \sigma_F^{(\ell_1\ell_2)}}{\frac38(1+\cos^2\theta)
  \sigma_U+\frac34\sin^2\theta\ \sigma_L+\frac34\cos\theta\ \sigma_F}
\end{equation}
where, according to Eq.~(\ref{sigma1}), the $\sigma_\alpha^{\{\ell_1\ell_2\}}$
are given by
\begin{eqnarray}
\sigma_\alpha^{\{\ell_1\ell_2\}}&=&g_{11}\sigma_\alpha^{1\,\{\ell_1\ell_2\}}
  +g_{12}\sigma_\alpha^{2\,\{\ell_1\ell_2\}},\qquad(\alpha=U,L)\nonumber\\
\sigma_F^{\{\ell_1\ell_2\}}&=&g_{43}\sigma_F^{3\,\{\ell_1\ell_2\}}
  +g_{44}\sigma_F^{4\,\{\ell_1\ell_2\}}.
\end{eqnarray}
As before the curly bracket notation $\{\ell_1\ell_2\}$ stands for either the
unpolarized or the longitudinal spin--spin correlation case. Because
$\sigma_\alpha^{j\,\{\ell_1\ell_2\}}$ and $H_\alpha^{j\,\{\ell_1\ell_2\}}$ are
proportional to each other, one can freely substitute
$H_\alpha^{j\,\{\ell_1\ell_2\}}$ for $\sigma_\alpha^{j\,\{\ell_1\ell_2\}}$ in
Eq.~(\ref{corrdef}).

In the forward and backward direction ($\cos\theta=\pm1$) one finds the
Born-term relation
\begin{equation}\label{pllborn1}
P^{\ell\ell}({\it Born};\cos\theta=\pm1)=-1
\end{equation}
regardless of the energy $\sqrt{q^2}$ which follows again from angular
momentum conservation since the spins of the quark and antiquark have to be
aligned in the forward and backward direction. Formally this can be seen by
substituting the Born-term relations~(\ref{born1}) into Eq.~(\ref{corrdef}).

Next we discuss the limiting values of the longitudinal spin--spin correlation
asymmetry at threshold $v\to 0$ and in the high energy limit $v\to 1$. At
threshold $v=0$ one finds\footnote{Note that in the threshold region QCD
binding effects modify the naive threshold results significantly.}
\begin{equation}\label{pllborn2}
\hspace{-2.0cm}\mbox{threshold:}\qquad\qquad
P^{\ell\ell}({\it Born};\cos\theta)=-\cos^2\theta
\end{equation}
which implies that the longitudinal spin--spin asymmetry vanishes at threshold
at $90^{\circ}$, i.e.\
$P^{\ell\ell}({\it Born};\cos\theta=0)_{\rm threshold}=0$. The angular
average is $\langle P^{\ell\ell}({\it Born};\cos\theta)_{\rm threshold}\rangle
=-1/3$ in agreement with Ref.~\cite{GKL2}. In the high energy limit
$m_q/\sqrt{q^2}\to 0$ where the quark and antiquark spins become completely
aligned, i.e. $\sigma_L=\sigma_L^{(\ell_1\ell_2)}=0$, one has
\begin{equation}\label{pllborn3}
\hspace{-3.0cm}\mbox{high energy limit:}\qquad\qquad
P^{\ell\ell}({\it Born};\cos\theta)=-1
\end{equation}
independent of the polar orientation.


\section{$O(\alpha_s)$ tree-graph contributions}


The $O(\alpha_s)$ spin-dependent hadronic three-body tensor
$H_{\mu\nu}(p_1,p_2,p_3,s_1^l,s_2^l)$ can easily be calculated from the
relevant Feynman diagrams. In order to calculate the longitudinal spin--spin
correlations one needs an explicit representation of the longitudinal spin
components of the quark and antiquark. They are given by
\begin{eqnarray}\label{spin1}
(s_1^\ell)^\mu&=&\frac{s_1^\ell}{\sqrt\xi}(\sqrt{(1-y)^2-\xi};0,0,1-y),\\
\label{spin2}
(s_2^\ell)^\mu&=&\frac{s_2^\ell}{\sqrt\xi}(\sqrt{(1-z)^2-\xi};
  (1-z)\sin\theta_{12},0,(1-z)\cos\theta_{12})
\end{eqnarray}
where $\theta_{12}$ is the polar angle between the quark and the
antiquark.\footnote{In the three-body case the quark and antiquark are no
longer back-to-back in general and thus the spins in the aligned spin
configurations $(\uparrow\downarrow)$ and $(\downarrow\uparrow)$ are no longer
parallel in general. However, in the $(t\bar t)$ case the top and antitop
quarks are quite ``stiff'' with respect to gluon radiation for the relevant
lower energies. The average opening angle is
$\langle\cos\theta_{12}\rangle=-0.996$ and
$\langle\cos\theta_{12}\rangle=-0.980$ at $\sqrt{q^2}=500\GeV$ and
$\sqrt{q^2}=1000\GeV$, respectively~\cite{Groote:1998xc}.} The polar angle is
given by ($\xi=4m_q^2/q^2=1-v^2, y=1-2p_1q/q^2, z=1-2p_2q/q^2$)
\begin{equation}
\cos\theta_{12}=\frac{yz+y+z-1+\xi}{\sqrt{(1-y)^2-\xi}\sqrt{(1-z)^2-\xi}}.
\end{equation}
In the two-body Born-term case, where $y=z=0$, one has $\cos\theta_{12}=-1$.
The polarization four-vectors~(\ref{spin1}) and~(\ref{spin2}) then simplify
to their expected two-body representations.

For the spin--spin dependent piece one obtains (with
$N:=\alpha_sN_cC_Fq^2/4\pi v$, $R_y:=\sqrt{(1-y)^2-\xi}$ and
$R_z:=\sqrt{(1-z)^2-\xi}$)
\begin{eqnarray}
H_U^{1\,(\ell_1\ell_2)}({\it tree})&=&\frac N{R_y^3R_z}
  \Bigg[-2(2-\xi)(24-23\xi+\xi^2)\nonumber\\&&
  +2(8-9\xi)(1-\xi)(2-\xi)\frac1y+4(30-24\xi+4\xi^2)y\nonumber\\[3pt]&&
  -2(44-26\xi+3\xi^2)y^2+8(5-\xi)y^3-8y^4\nonumber\\[3pt]&&
  +2\xi(2-\xi)(1-\xi)^2\left(\frac1{y^2}+\frac1{z^2}\right)
  -6\xi(1-\xi)(2-\xi)\frac{y}{z^2}\nonumber\\&&
  +4\xi(2-\xi)^2\frac{y^2}{z^2}-4\xi(3-2\xi)\frac{y^3}{z^2}
  +2\xi(2-\xi)\frac{y^4}{z^2}\nonumber\\&&
  +2(16-11\xi)(1-\xi)(2-\xi)\frac1z-4(1-\xi)^2(2-\xi)^2\frac1{yz}\nonumber\\&&
  -2(2-\xi)(26-27\xi+3\xi^2)\frac yz+8(11-13\xi+4\xi^2)\frac{y^2}z\nonumber\\&&
  -4(1-\xi)(10-\xi)\frac{y^3}z+8(1-\xi)\frac{y^4}z\nonumber\\&&
  +8(5-3\xi)z-2\xi(1-\xi)(2-\xi)\frac{z}{y^2}\nonumber\\&&
  -2(2-\xi)(6-9\xi+\xi^2)\frac zy-4(6-3\xi+2\xi^2)yz\nonumber\\&&
  +8(1-2\xi)y^2z-4(2-2\xi+\xi^2)z^2
  +2\xi(1-\xi)(2-\xi)\frac{z^2}{y^2}\nonumber\\&&
  +4(2-\xi)(1-2\xi)\frac{z^2}y+8yz^2-16\xi yz^2-8y^2z^2\Bigg],\\
H_U^{2\,(\ell_1\ell_2)}({\it tree})&=&-\frac{\xi N}{R_y^3R_z}
  \Bigg[40-36\xi-2\xi(1-\xi)^2\frac1{y^2}-2(8-9\xi)(1-\xi)\frac1y\nonumber\\&&
  -2(12-\xi)y-2(4-\xi)y^2+8y^3-2\xi(1-\xi)^2\frac1{z^2}\nonumber\\&&
  +6\xi(1-\xi)\frac{y}{z^2}-2\xi(3-\xi)\frac{y^2}{z^2}
  +\frac{2\xi y^3}{z^2}\nonumber\\&&
  -2(16-11\xi)(1-\xi)\frac1z+4(1-\xi)^2(2-\xi)\frac1{yz}\nonumber\\&&
  +2(24-23\xi+\xi^2)\frac yz-2(16-5\xi)\frac{y^2}z\nonumber\\&&
  +2(4+\xi)\frac{y^3}z-2(16-3\xi)z+2\xi(1-\xi)\frac{z}{y^2}\nonumber\\&&
  +2(8-11\xi+\xi^2)\frac zy+2(8-\xi)yz\nonumber\\&&
  +2(8-\xi)z^2-2\xi(1-\xi)\frac{z^2}{y^2}-4(2-3\xi)\frac{z^2}y-8yz^2\Bigg],\\
H_L^{1\,(\ell_1\ell_2)}({\it tree})&=&\frac N{R_y^3R_z}
  \Bigg[-16+56\xi-43\xi^2+5\xi^3-\xi^2(1-\xi)^2\frac1{y^2}\nonumber\\&&
  -\xi(8-9\xi)(1-\xi)\frac1y+(32-70\xi+23\xi^2)y\nonumber\\&&
  -2(8-9\xi)y^2-6\xi y^3+2\xi y^4-\xi^2(1-\xi)^2\frac1{z^2}\nonumber\\&&
  +3\xi^2(1-\xi)\frac{y}{z^2}-2\xi^2(2-\xi)\frac{y^2}{z^2}
  +\frac{\xi^2y^3}{z^2}-\frac{\xi^2y^4}{z^2}\nonumber\\&&
  -\xi(16-11\xi)(1-\xi)\frac1z+2\xi(1-\xi)^2(2-\xi)\frac1{yz}\nonumber\\&&
  +\xi(30-33\xi+5\xi^2)\frac yz-9\xi(2-\xi)\frac{y^2}z\nonumber\\&&
  +\xi(2-\xi)\frac{y^3}z-\frac{2\xi y^4}z+(32-66\xi+23\xi^2)z\nonumber\\&&
  +\xi^2(1-\xi)\frac{z}{y^2}+\xi(10-15\xi+3\xi^2)\frac zy
  -(48-38\xi+\xi^2)yz\nonumber\\&&
  +2(8-\xi)y^2z+4\xi y^3z-(16-26\xi+\xi^2)z^2
  -\xi^2(1-\xi)\frac{z^2}{y^2}\nonumber\\&&
  -2\xi(3-4\xi)\frac{z^2}y+2(8+\xi)yz^2+2\xi y^2z^2\Bigg],\\
H_L^{2\,(\ell_1\ell_2)}({\it tree})&=&\frac{\xi N}{R_y^3R_z}
  \Bigg[24-25\xi+3\xi^2-\xi(1-\xi)^2\frac1{y^2}\nonumber\\&&
  -(8-9\xi)(1-\xi)\frac1y-(22-9\xi)y+2y^2+6y^3-2y^4\nonumber\\&&
  -\xi(1-\xi)^2\frac1{z^2}+3\xi(1-\xi)\frac{y}{z^2}
  -\frac{2\xi y^2}{z^2}-\frac{\xi y^3}{z^2}+\frac{\xi y^4}{z^2}\nonumber\\&&
  -(16-11\xi)(1-\xi)\frac1z+2(1-\xi)^2(2-\xi)\frac1{yz}\nonumber\\&&
  +(22-21\xi+\xi^2)\frac yz
  -(10-3\xi)\frac{y^2}z-(2-\xi)\frac{y^3}z+\frac{2y^4}z\nonumber\\&&
  -13(2-\xi)z+\xi(1-\xi)\frac{z}{y^2}+(10-15\xi+3\xi^2)\frac zy\nonumber\\&&
  +(18+\xi)yz+2y^2z-4y^3z+(10+\xi)z^2\nonumber\\&&
  -\xi(1-\xi)\frac{z^2}{y^2}-2(3-4\xi)\frac{z^2}y-2yz^2-2y^2z^2\Bigg],\\
H_F^{4\,(\ell_1\ell_2)}({\it tree})&=&\frac{2N}{R_y^2R_z}
  \Bigg[-32+26\xi+\xi^2+2\xi(1-\xi)^2\frac1{y^2}
  +16(1-\xi)^2\frac1y\nonumber\\&&
  +4(7-\xi)y-4(4-\xi)y^2+4y^3+2\xi(1-\xi)^2\frac1{z^2}\nonumber\\&&
  -4\xi(1-\xi)\frac{y}{z^2}
  +\xi(4-3\xi)\frac{y^2}{z^2}-\xi(2-\xi)\frac{y^3}{z^2}\nonumber\\&&
  +6(4-3\xi)(1-\xi)\frac1z
  -4(1-\xi)^2(2-\xi)\frac1{yz}-4yz^2\nonumber\\&&
  -(28-32\xi+3\xi^2)\frac yz+2(1-\xi)(8-\xi)\frac{y^2}z-4(1-\xi)\frac{y^3}z
  \nonumber\\&&
  +2(4-\xi)(1+\xi)z-2\xi(1-\xi)\frac{z}{y^2}-(12-16\xi+\xi^2)\frac zy
  \nonumber\\&&
  +4(1-\xi)yz-4\xi z^2+2\xi(1-\xi)\frac{z^2}{y^2}+(4-6\xi-\xi^2)\frac{z^2}y
  \Bigg].\qquad
\end{eqnarray}
The unpolarized helicity structure functions are needed for normalization
purposes. They read (see e.g.\ Refs.~\cite{GKT2})
\begin{eqnarray}
H_U^1({\it tree})&=&\frac N{R_y^2}\Bigg[16-4\xi-4\xi^2-\frac{4\xi}{y^2}
  +\frac{6\xi^2}{y^2}-\frac{2\xi^3}{y^2}-\frac{16}y+\frac{32\xi}y
  -\frac{12\xi^2}y\nonumber\\&&
  -8\xi y-\frac{4\xi}{z^2}+\frac{6\xi^2}{z^2}-\frac{2\xi^3}{z^2}
  +\frac{8\xi y}{z^2}-\frac{4\xi^2y}{z^2}-\frac{4\xi y^2}{z^2}-\frac{48}{z}
  +\frac{56\xi}z\nonumber\\&&
  -\frac{16\xi^2}z+\frac{16}{yz}-\frac{32\xi}{yz}+\frac{20\xi^2}{yz}
  -\frac{4\xi^3}{yz}+\frac{56y}{z}-\frac{28\xi y}{z}-\frac{2\xi^2y}z
  \nonumber\\&&
  -\frac{32y^2}z+\frac{8y^3}z-8\xi z+\frac{8z}y-\frac{4\xi z}y
  -\frac{2\xi^2z}y+8yz\Bigg],\\
H_U^2({\it tree})&=&\frac{\xi N}{R_y^2}\Bigg [8+2\xi-\frac{2\xi}{y^2}
  +\frac{2\xi^2}{y^2}-\frac8y+\frac{12\xi}y-\frac{2\xi}{z^2}
  +\frac{2\xi^2}{z^2}\nonumber\\&&
  +\frac{4\xi y}{z^2}-\frac{2\xi y^2}{z^2}-\frac{24}z+\frac{16\xi}z
  +\frac8{yz}-\frac{12\xi}{yz}+\frac{4\xi^2}{yz}\nonumber\\&&
  +\frac{24y}z-\frac{2\xi y}z-\frac{8y^2}{z}+8z+\frac{2\xi z}y\Bigg],\\
H_L^1({\it tree})&=&\frac N{R_y^2}\Bigg [16-4\xi+\xi^2-\frac{\xi^2}{y^2}
  +\frac{\xi^3}{y^2}-\frac{4\xi}y+\frac{6\xi^2}y-16y+4\xi y+4\xi y^2
  \nonumber\\&&
  -\frac{\xi^2}{z^2}+\frac{\xi^3}{z^2}+\frac{2\xi^2y}{z^2}
  +\frac{\xi^2y^2}{z^2}-\frac{12\xi}z+\frac{8\xi^2}z
  +\frac{4\xi}{yz}-\frac{6\xi^2}{yz}+\frac{2\xi^3}{yz}\nonumber\\&&
  +\frac{2\xi y}z+\frac{2\xi^2y}z+\frac{2\xi y^3}z-16z+4\xi z
  -\frac{2\xi z}y+2\xi yz\Bigg],\\
H_L^2({\it tree})&=&\frac{\xi N}{R_y^2}\Bigg[12-\xi-\frac\xi{y^2}
  +\frac{\xi^2}{y^2}-\frac4y+\frac{6\xi}y-4y-4y^2\nonumber\\&&
  -\frac\xi{z^2}+\frac{\xi^2}{z^2}+\frac{2\xi y}{z^2}
  -\frac{\xi y^2}{z^2}-\frac{12}z+\frac{8\xi}z+\frac4{yz}
  -\frac{6\xi}{yz}+\frac{2\xi^2}{yz}\nonumber\\&&
  +\frac{10y}z-\frac{2\xi y}z-\frac{2y^3}z-4z-\frac{2z}y-2yz\Bigg],\\
H_F^4({\it tree})&=&\frac{4N}{R_y}\Bigg[-\frac\xi{y^2}+\frac{\xi^2}{y^2}+
  -\frac4y+\frac{5\xi}y-\frac\xi{z^2}+\frac{\xi^2}{z^2}+\frac{\xi y}{z^2}
  -\frac8z+\frac{6\xi}z\nonumber\\&&
  +\frac4{yz}-\frac{6\xi}{yz}+\frac{2\xi^2}{yz}+\frac{6y}z-\frac{2y^2}z
  +2z+\frac{2z}y\Bigg].
\end{eqnarray}


\section{One-loop contributions and $O(\alpha_s)$ corrections}


The one-loop contributions to the unpolarized helicity structure functions
can be taken from Refs.~\cite{KPT,GKT2}. Let us first list the contributions
from the real part of the one-loop amplitude. For the unpolarized case one has
\begin{eqnarray}\label{uloop}
H_U^1({\it loop\/})&=&4N_cq^2(\real A+v^2\real C),\nonumber\\
H_U^2({\it loop\/})&=&4N_cq^2(\real A-v^2\real C),\nonumber\\
H_L^1({\it loop\/})&=&H_L^2({\it loop\/})\ =\ 2N_cq^2(\xi\real A+v^2\real B),
  \nonumber\\
H_F^4({\it loop\/})&=&4N_cq^2v(\real A+\real C).
\end{eqnarray}
For the longitudinal spin--spin correlation components one obtains
\begin{eqnarray}\label{ploop}
H_U^{1\,(\ell_1\ell_2)}({\it loop\/})&=&-4N_cq^2(\real A+v^2\real C),
  \nonumber\\
H_U^{2\,(\ell_1\ell_2)}({\it loop\/})&=&-4N_cq^2(\real A-v^2\real C),
  \nonumber\\
H_L^{1\,(\ell_1\ell_2)}({\it loop\/})&=&H_L^{2\,(\ell_1\ell_2)}({\it loop\/})
  \ =\ 2N_cq^2(\xi\real A+v^2\real B),\nonumber\\
H_F^{4\,(\ell_1\ell_2)}({\it loop\/})&=&-4N_cq^2v(\real A+\real C).
\end{eqnarray}
Since the one-loop contributions refer to two-body final states they also
satisfy the two-body relations~(\ref{born1}).

There are also contributions coming from the imaginary part of the vertex
correction which multiplies the imaginary part of the Breit--Wigner function
of the $Z$ resonance as indicated in the electroweak model parameters
$g_{43}$ and $g_{13}$ in Eqs.~(\ref{gdiag}) and~(\ref{gnondiag}). The relevant
hadron tensor component results from the $VA$ interference term in the $F$
projection. Although these contributions are rather small (especially when
one is far away from the $Z$ resonance) they are included in our numerical
results. The imaginary part contributions are given by
\begin{eqnarray}
H_F^3({\it loop\/})&=&-8N_cq^2v\imag B,\nonumber\\
H_F^{3\,(\ell_1\ell_2)}({\it loop\/})&=&-8N_cq^2v\imag B.
\end{eqnarray}

The one-loop form factors $A$, $B$, and $C$ appearing in Eqs.~(\ref{uloop})
and~(\ref{ploop}) read ($C_{F}=4/3$)~\cite{GKT2}
\begin{eqnarray}
\real A&=&-\frac{\alpha_sC_F}{4\pi}\Bigg[\left(2+\frac{1+v^2}v
  \ln\pfrac{1-v}{1+v}\right)\ln\pfrac{\Lambda q^2}{m^2}+3v\ln\pfrac{1-v}{1+v}
  +4\nonumber\\&&
  +\frac{1+v^2}v\left(\Li_2\pfrac{2v}{1+v}+\frac14\ln^2\pfrac{1-v}{1+v}
  -\frac{\pi^2}2\right)\Bigg],\nonumber\\
\real B&=&\frac{\alpha_sC_F}{4\pi}\ \frac{1-v^2}v\ln\pfrac{1-v}{1+v},\qquad
\real C\ =\ \real A-2\real B,\nonumber\\
\imag B&=&\frac{\alpha_sC_F}{4\pi}\ \frac{1-v^2}v\pi.
\end{eqnarray}
The one-loop contributions have been infrared regularized by introducing a
(small) gluon mass $m_g=\sqrt{\Lambda q^2}$ following Refs.~\cite{GKT2}.

What remains to be done is to perform the phase space integrations over the
tree-graph contributions listed in Sec.~4. One first integrates over $z$
and then over $y$. The relevant integration limits incorporating the auxiliary
gluon mass are
\begin{equation}
z_\pm(y)=\frac{2y}{4y+\xi}\left\{1-y-\frac12\xi+\Lambda+\frac\Lambda y
\pm\frac1y\sqrt{(y-\Lambda)^2-\Lambda\xi}\sqrt{(1-y)^2-\xi}\right\}
\end{equation}
and
\begin{equation}
y_-=\sqrt{\Lambda\xi}+\Lambda,\qquad y_+=1-\sqrt\xi.
\end{equation}

The introduction of a (small) gluon mass distorts phase space away from the
infrared singularity at $y=z=0$. The infrared singularities in the tree-graph
and one-loop contributions cancel and one remains with finite remainders. It
is quite clear that the finite result in the sum is independent of the
specific regularization procedure. Finally, adding in the above one-loop
contributions, one obtains the $O(\alpha_s)$ corrections (loop + tree). For the
sake of completeness we include in our results also the unpolarized hadron
tensor components which are needed for the normalization of the longitudinal
spin--spin asymmetry. One obtains
\begin{eqnarray}
H_U^1(\alpha_s)&=&N\Bigg[2(2+7\xi)v-8(2-\xi)vt_{10}-16(2-\xi)vt_{12}
  \nonumber\\&&
  +(48-48\xi+7\xi^2)t_3+2\sqrt\xi(1-\sqrt\xi)(2+4\sqrt\xi-3\xi)t_4\nonumber\\&&
  -2\xi(2+3\xi)t_5-4(2-\xi)^2(t_8-t_9)\Bigg],\\
H_U^2(\alpha_s)&=&\xi N\Bigg[12v-8vt_{10}-16vt_{12}+2(6-\xi)t_3
  +2\sqrt\xi(1-\sqrt\xi)t_4\nonumber\\&&
  +2\xi t_5-4(2-\xi)(t_8-t_9)\Bigg],\\
H_L^1(\alpha_s)&=&N\Bigg[\left(8-23\xi+\frac{3\xi^2}2\right)v-4\xi vt_{10}
  -8\xi vt_{12}+\nonumber\\&&
  \xi\left(22-8\xi+\frac{3\xi^2}4\right)t_3
  -2\sqrt\xi(1-\sqrt\xi)(2+4\sqrt\xi-3\xi)t_4\nonumber\\&&
  +2\xi(2+3\xi)t_5-2\xi(2-\xi)(t_8-t_9)\Bigg],\\
H_L^2(\alpha_s)&=&\xi N\Bigg[\frac32(10-\xi)v-4vt_{10}-8vt_{12}
  +\left(6-4\xi-\frac{3\xi^2}4\right)t_3\nonumber\\&&
  -2\sqrt\xi(1-\sqrt\xi)t_4-2\xi t_5-2(2-\xi)(t_8-t_9)\Bigg],\\
H_F^4(\alpha_s)&=&N\Bigg[-16\sqrt\xi(1-\sqrt\xi)-16v^2(t_{10}+t_{11})
  -16(t_1-t_2)\nonumber\\&&
  +8(2-3\xi)vt_3-4(4-5\xi)t_6+8(2-\xi)v(t_7-t_8)\Bigg]
\end{eqnarray}
and
\begin{eqnarray}
H_U^{1\,(\ell_1\ell_2)}(\alpha_s)&=&N\Bigg[8+13\xi-\sqrt\xi(4+25\xi)
  -\frac1{2v}(10-7\xi-19\xi^2)\nonumber\\&&
  -v\xi\left(t_1+t_{12}-\frac{t_{11}}2-\ln 4\right)+\nonumber\\&&
  +2(2-\xi)\left(\frac{2-\xi}2(t_8-t_{16})+v(t_{10}+2t_{12})\right)
  \nonumber\\&&
  -\frac12(8+2\xi+3\xi^2)t_{14}+\frac1{v^3}(8-15\xi+13\xi^2-4\xi^3)t_{15}
  \nonumber\\&&
  -\frac1{2v^2}(16-36\xi+55\xi^2-11\xi^3-4v^3\xi)t_{13}+\nonumber\\&&
  -\frac1{v^2}\left(4-3\xi+\xi^2+v^3(8-3\xi)\right)t_3\Bigg],\\
H_U^{2\,(\ell_1\ell_2)}(\alpha_s)&=&\xi N\Bigg[6-5\sqrt\xi-\frac1v
  -v\left(t_1+t_{12}-\frac{t_{11}}2-\ln 4\right)-\frac12(8+\xi)t_{14}
  \nonumber\\&&
  +\frac1{2v^2}(4-\xi+4v^3)t_{13}
  +2\left(\frac{2-\xi}2(t_8-t_{16})+v(t_{10}+2t_{12})\right)\nonumber\\&&
  +\frac1{4v^3}(2-3\xi)(8-7\xi)t_{15}
  +\frac1{4v^2}\left(\xi(9-8\xi)-20v^3\right)t_3\Bigg],\\
H_L^{1\,(\ell_1\ell_2)}(\alpha_s)&=&N\Bigg[-\frac34(24+22\xi-\xi^2)+
  \frac14\sqrt\xi(48+112\xi+3\xi^2)\nonumber\\&&
  +\frac1{8v}(128-106\xi-81\xi^2+3\xi^3)
  +v\xi\left(t_1+t_{12}-\frac{t_{11}}2-\ln 4\right)\nonumber\\&&
  -\xi\left(\frac{2-\xi}2(t_8-t_{16})+v(t_{10}+2t_{12})\right)
  +\xi(6+\xi)t_{14}\nonumber\\&&
 -\frac1{8v^2}(64-32\xi-164\xi^2+51\xi^3-3\xi^4+16v^3\xi)t_{13}\\&&
  -\frac\xi{4v^3}(28-24\xi+3\xi^2)t_{15}
  +\frac\xi{4v^2}(30-27\xi+4\xi^2+12v^3)t_3\Bigg],\nonumber\\
H_L^{2\,(\ell_1\ell_2)}(\alpha_s)&=&\xi N\Bigg[\frac14(34-3\xi)
  -\frac34\sqrt\xi(12+\xi)-\frac1{8v}(46-65\xi+3\xi^2)\nonumber\\&&
  -(1+\xi)t_{14}-\left(\frac{2-\xi}2(t_8-t_{16})+v(t_{10}+2t_{12})\right)
  \nonumber\\&&
  +\frac1{8v^2}(80-136\xi+35\xi^2-3\xi^3)t_{13}\nonumber\\&&
  -\frac1{2v^3}(4-9\xi+4\xi^2)t_{15}
  +\frac1{2v^2}(1-4\xi+2\xi^2+4v^3)t_3\Bigg],\\
H_F^{3\,(\ell_1\ell_2)}(\alpha_s)&=&N\Bigg[2\pi v\xi \Bigg],\\
H_F^{4\,(\ell_1\ell_2)}(\alpha_s)&=&N\Bigg[2\sqrt\xi+4v^2
  +\frac{2\xi}{v^2}(5-3\xi)(t_1-\ln 4)\nonumber\\&&
  +2v^2t_{11}-(4+\xi)t_{14}+2v(2-\xi)(t_8-t_{16}+t_{22})\nonumber\\&&
  +\frac1{4v^2}(16-32\xi+18\xi^2-\xi^3)t_{21}\nonumber\\&&
  +\frac\xi{v^3}(5-8\xi+4\xi^2)(t_{20}-t_{19})\nonumber\\&&
  +\frac{\sqrt\xi}{2v^3}(8-11\xi+3\xi^2+2\xi^3)(t_{20}+t_{19})\nonumber\\&&
  +\frac\xi{4v^2}(20-22\xi+\xi^2)(t_{18}-t_{17})\nonumber\\&&
  +\frac{\sqrt\xi}{4v^2}(16-10\xi-5\xi^2)(t_{18}+t_{17})\nonumber\\&&
  -\frac{\sqrt\xi}{2v^2}(2+7\xi-5\xi^2)(t_{12}-t_{10})\nonumber\\&&
  +\frac1{v^2}(2-\xi)(1+\xi)(t_{12}+t_{10})\nonumber\\&&
  +\frac1{v^3}(4-5\xi+2\xi^3)t_{15}
  +\frac1{2v^2}(8-18\xi+\xi^2+5\xi^3)t_{13}\nonumber\\&&
  -\frac1{4v^2}\left(8-18\xi+\xi^2+5\xi^3+2v(8-14\xi+7\xi^2)\right)t_3\Bigg]
\end{eqnarray}
The unpolarized hadron tensor components $H_U^1(\alpha_s)$, $H_U^2(\alpha_s)$,
$H_L^1(\alpha_s)$, $H_L^2(\alpha_s)$, and $H_F^4(\alpha_s)$ including the
$O(\alpha_s)$ integrated rate functions $t_i$ ($i=1,2,3,6,7,8,9,10,12$) have
been calculated and listed before in Ref.~\cite{GKT2}. The integrated rate
functions $t_i$ ($i=13,14,15,16$) appear in the evaluation of the longitudinal
spin--spin component $H_{U+L}^{4\,(\ell_1\ell_2)}$ and have been listed in
Ref.~\cite{GKL2}. In addition to the decay rate functions calculated in
Ref.~\cite{GKL2,GKT2} the spin--spin polar dependence contributions bring in a
set of new decay rate functions $t_i$ ($i=17,18,19,20,21,22$). The additional
set of decay rate functions needed in the present application is given by
\begin{eqnarray}
t_{17}&=&\frac52\ln\pfrac{1+v}{2-\sqrt\xi}\ln\pfrac{2-\sqrt\xi}2
  +\frac12\ln\pfrac{1-v}{2-\sqrt\xi}\ln\pfrac{1+v}2\nonumber \\&&
  +\Li_2(1)-\Li_2\left(\sqrt{\frac{1-v}{1+v}}\right)
  -2\Li_2\pfrac\xi{(2-\sqrt\xi)^2}\nonumber\\&&
  +2\Li_2\pfrac{\sqrt\xi}{2-\sqrt\xi}
  +\Li_2\pfrac{(1-v)^2}{\sqrt\xi(2-\sqrt\xi)}-\Li_2\pfrac{2-\sqrt\xi}{1+v},\\
t_{18}&=&\ln\pfrac{1+v}{2-\sqrt\xi}\ln\pfrac2{\sqrt\xi}
  +\ln\pfrac{\sqrt\xi(1+v)}{(2-\sqrt\xi)^2}\ln\pfrac2{2+\sqrt\xi}\nonumber\\&&
  +\Li_2(-1)+\Li_2\pfrac{2-\sqrt\xi}{2+\sqrt\xi}
  -\Li_2\left(-\frac{2-\sqrt\xi}{2+\sqrt\xi}\right)\nonumber\\&&
  -2\Li_2\pfrac{-\sqrt\xi}{2-\sqrt\xi}+2\Li_2\pfrac{-\xi}{4-\xi}
  -\Li_2\pfrac{-(1-v)^2}{\sqrt\xi(2+\sqrt\xi)}\nonumber\\&&
  -\Li_2\pfrac{1+v}{2+\sqrt\xi}+\Li_2\left(-\sqrt{\frac{1-v}{1+v}}\right),\\
t_{19}&=&\ln\pfrac{1+v}{1-v}\ln\pfrac{1+v}{2-\sqrt\xi}
  +\ln^2\pfrac{1+v}{2-\sqrt\xi}\nonumber\\&&
  +2\Li_2\pfrac{1-v}{2-\sqrt\xi}+2\Li_2\pfrac{2-\sqrt\xi}{1+v}-4\Li_2(1),\\
t_{20}&=&\ln\pfrac{1+v}{1-v}\ln\pfrac{2-\sqrt\xi}{1+v}+2\Li_2(-w)-2\Li_2(w)
  \nonumber\\&&
  -2\Li_2\left(-\frac{1-v}{2-\sqrt\xi}w\right)
  +2\Li_2\left(\frac{1+v}{2-\sqrt\xi}w\right),\\
t_{21}&=&3\left(\frac12\ln^2\pfrac{1+v}{1-v}
  +\ln\left(\frac4\xi\sqrt{\frac{1-v}{1+v}}\right)
  \ln\pfrac{\sqrt\xi}{2-\sqrt\xi}\right)\nonumber\\&&
  +3\Bigg(\Li_2\pfrac{2-\sqrt\xi}{1+v}
  -\Li_2\left(\frac{2-\sqrt\xi}{1+v}\sqrt{\frac{1-v}{1+v}}\right)\nonumber\\&&
  +\Li_2\left(\frac{1-v}{2-\sqrt\xi}\sqrt{\frac{1-v}{1+v}}\right)
  -\Li_2\pfrac{1-v}{2-\sqrt\xi}\Bigg)\nonumber\\&&
  +\Li_2\pfrac{4v}{(1+v)^2}+\Li_2\pfrac{-2v}{1-v}-\Li_2\pfrac{2v}{1+v},\\
t_{22}&=&2\ln\pfrac{1+v}{2-\sqrt\xi}
  \ln\pfrac{2(1+\sqrt\xi)(2-\sqrt\xi)}{(1+v)^2}\nonumber\\&&
  +4\Li_2\pfrac{\sqrt\xi-1+v}{2v}
  -4\Li_2\pfrac{(1-v)(\sqrt\xi-1+v)}{2v(2-\sqrt\xi)}
\end{eqnarray}
where
\begin{equation}
w=\sqrt{\frac{1-\sqrt\xi}{1+\sqrt\xi}}.
\end{equation}
We have checked that the sum of $H_U^{1,2\,(\ell_1\ell_2)}(\alpha_s)$
and $H_L^{1,2\,(\ell_1\ell_2)}(\alpha_s)$ agrees with the form 
$H_{U+L}^{1,2\,(\ell_1\ell_2)}(\alpha_s)$ calculated in Ref.~\cite{GKL2}.


\section{Massless QCD and the zero-mass limit of QCD}


Before we turn to the numerical evaluation of the longitudinal spin--spin
correlations, we would like to discuss the $m_q\to 0$ limit of our analytical
results with the aim to compare them to the corresponding results obtained in
massless QCD ($m_q=0$). As is well-known by now, the $m_q\to 0$ limit in
particular of the spin-flip contribution does not coincide with that of
$m_{q}=0$ QCD where there is no spin-flip~\cite{LeeNau,Smi,Falk,KPT,GKT2}.

The $m_q=0$ expressions can be calculated in dimensional regularization as
described in Ref.~\cite{GKT2}. For the $O(\alpha_s)$ contributions one obtains
\begin{eqnarray}
H_U^1(\alpha_s)\ =\ 4N_cq^2\frac{\alpha_sC_F}{4\pi},&&
H_U^{1\,(\ell_1\ell_2)}(\alpha_s)\ =\ -4N_cq^2\frac{\alpha_sC_F}{4\pi},
  \nonumber\\
H_L^1(\alpha_s)\ =\ 8N_cq^2\frac{\alpha_sC_F}{4\pi},&&
H_L^{1\,(\ell_1\ell_2)}(\alpha_s)\ =\ -8N_cq^2\frac{\alpha_sC_F}{4\pi},
  \nonumber\\
H_F^4(\alpha_s)\ =\ 0,&&
H_F^{4\,(\ell_1\ell_2)}(\alpha_s)\ =\ 0.
\end{eqnarray}
The $O(\alpha_s)$ $m_q=0$ single-spin functions
$H_\alpha^{j\,(\ell_{1,2})}(\alpha_s)$ are given in Ref.~\cite{GKT2} or can
be read off from Eq.~(\ref{singlespin2}).

Let us now consider the $m_q\to 0$ limit of the integrated decay rate
functions $t_i$ discussed in Sec.~5. One obtains\\
{\renewcommand{\arraystretch}{2.2}
\begin{tabular}{lll}
$\displaystyle t_1\to\ln 4-\frac32\ln\pfrac4\xi$&
$\displaystyle t_2\to\ln 4-\frac12\ln\pfrac4\xi$&\kern-9pt
$\displaystyle t_3\to\ln\pfrac4\xi$\quad
$\displaystyle t_4\to\frac{\pi^2}2$\\
$\displaystyle t_5\to\frac{\pi^2}6-\frac14\ln^2\pfrac4\xi$&\kern-4pt
$\displaystyle t_6\to\frac{\pi^2}6+\frac14\ln^2\pfrac4\xi$&\kern-9pt
$\displaystyle t_7\to-\frac{\pi^2}2-\frac14\ln^2\pfrac4\xi$\\
$\displaystyle t_8\to-\frac{2\pi^2}3-\frac12\ln^2\pfrac4\xi$&
$\displaystyle t_9\to-\frac{2\pi^2}3-\frac12\ln^2\pfrac4\xi$&
$\displaystyle t_{10}\to\ln\pfrac4\xi$\\
$\displaystyle t_{11}\to\ln\pfrac4\xi$&
$\displaystyle t_{12}\to\ln\pfrac4\xi$&
$\displaystyle t_{13}\to0$\\
$\displaystyle t_{14}\to\frac{\pi^2}6$&
$\displaystyle t_{15}\to\frac{\pi^2}3+\frac14\ln^2\pfrac4\xi$&
$\displaystyle t_{16}\to-\frac{\pi^2}6$\\
$\displaystyle t_{17}\to0$&
$\displaystyle t_{18}\to0$&
$\displaystyle t_{19}\to-\frac{\pi^2}3$\\
$\displaystyle t_{20}\to-\frac{\pi^2}6$&
$\displaystyle t_{21}\to\frac{\pi^2}3+\frac14\ln^2\pfrac4\xi$&
$\displaystyle t_{22}\to0$
\end{tabular}}\vspace{-12pt}
\begin{equation}\end{equation}
Using these limiting expressions one obtains for the $O(\alpha_s)$ $m_q\to 0$
unpolarized structure functions (loop + tree)
\begin{equation}
\label{helimit1}
H_U^1(\alpha_s)=4N_cq^2\frac{\alpha_sC_F}{4\pi},\qquad
H_L^1(\alpha_s)=8N_cq^2\frac{\alpha_sC_F}{4\pi},\qquad
H_F^4(\alpha_s)=0
\end{equation}
and, for the longitudinal spin--spin correlation functions,
\begin{eqnarray}
\label{helimit2}
H_U^{1(\ell_1,\ell_2)}(\alpha_s)&=&12N_cq^2\frac{\alpha_sC_F}{4\pi}
  \ =\ (-4+[16])N_cq^2\frac{\alpha_sC_F}{4\pi},\nonumber\\
H_L^{1(\ell_1,\ell_2)}(\alpha_s)&=&-8N_cq^2\frac{\alpha_sC_F}{4\pi},\nonumber\\
H_F^{4(\ell_1,\ell_2)}(\alpha_s)&=&16N_cq^2\frac{\alpha_sC_F}{4\pi}
  \ =\ [16]N_cq^2\frac{\alpha_sC_F}{4\pi}.
\end{eqnarray}
With the square bracket notation we indicate the difference between the
$O(\alpha_s)$ $m_q\to 0$ and $m_q=0$ results which we call anomalous
terms.\footnote{We use the phrase ``anomalous flip contribution'' since the
same anomalous flip contribution contributes to the absorptive part of the
$VVA$ triangle diagram in the zero-mass limit of QCD which in turn can be
related to the well-known axial anomaly via a dispersion relation
approach~\cite{dz71}.} Note that the current-current structures for $j=2,3$
are zero in the limit $m_q\to 0$ (and also for $m_q=0$).

For completeness we also list the $O(\alpha_s)$ $m_q\to 0$ single-spin
functions using the same square bracket notation. They read~\cite{GKT2}
\begin{eqnarray}\label{singlespin2}
H_U^{4\,(\ell_{1,2})}(\alpha_s)&=&\pm(4-[8])N_cq^2\frac{\alpha_sC_F}{4\pi},
\nonumber\\
H_L^{4\,(\ell_{1,2})}(\alpha_s)&=&\pm(8-[0])N_cq^2\frac{\alpha_sC_F}{4\pi},
\nonumber\\
H_F^{1\,(\ell_{1,2})}(\alpha_s)&=&\pm(0-[8])N_cq^2\frac{\alpha_sC_F}{4\pi}.
\end{eqnarray}

We finally collect all our $m_q\to 0$ results, including the Born-term
contributions Eqs.~(\ref{uborn}) and~(\ref{pborn}), where we make use of the
representation Eq.~(\ref{eqnkd1}). Again we split off the anomalous
contributions using the square bracket notation. One obtains ($C_F=4/3$ is
made explicit here)
\begin{eqnarray}\label{anomalies}
H^1_U(s_1^\ell,s_2^\ell)
  &=&\frac14\Big(H^1_U+H^{1\,(\ell_1\ell_2)}_Us_1^\ell s_2^\ell\Big)\nonumber\\
  &=&N_cq^2\left((1-s_1^\ell s_2^\ell)
  \left(1+\frac13\times\frac{\alpha_s}\pi\right)
  +\left[\frac43\times\frac{\alpha_s}\pi s_1^\ell s_2^\ell\right]\right),
  \nonumber\\
H^1_L(s_1^\ell,s_2^\ell)
  &=&\frac14\Big(H^1_L+H^{1\,(\ell_1\ell_2)}_Ls_1^\ell s_2^\ell\Big)\nonumber\\
  &=&N_cq^2(1-s_1^\ell s_2^\ell)\left(0+\frac23\times\frac{\alpha_s}\pi
  +[0]\right),\nonumber\\
H^1_F(s_1^\ell,s_2^\ell)
  &=&\frac14\Big(H^{1(\ell_1)}_Fs_1^\ell+H^{1(\ell_2)}_Fs_2^\ell\Big)
  \nonumber\\
  &=&N_cq^2(s_1^\ell-s_2^\ell)\left(1+0\times\frac{\alpha_s}\pi
  -\left[\frac23\times\frac{\alpha_s}\pi\right]\right),\nonumber\\
H^4_U(s_1^\ell,s_2^\ell)
  &=&\frac14\Big(H^{4(\ell_1)}_Us_1^\ell+H^{4(\ell_2)}_Us_2^\ell\Big)
  \nonumber\\
  &=&N_cq^2(s_1^\ell-s_2^\ell)\left(1+\frac13\times\frac{\alpha_s}\pi
  -\left[\frac23\times\frac{\alpha_s}\pi\right]\right),\nonumber\\
H^4_L(s_1^\ell,s_2^\ell)
  &=&\frac14\Big(H^{4(\ell_1)}_Ls_1^\ell+H^{4(\ell_2)}_Ls_2^\ell\Big)
  \nonumber\\
  &=&N_cq^2(s_1^\ell-s_2^\ell)\left(0+\frac23\times\frac{\alpha_s}\pi
  +[0]\right),\nonumber\\
H^4_F(s_1^\ell,s_2^\ell)
  &=&\frac14\Big(H^4_F+H^{4\,(\ell_1\ell_2)}_Fs_1^\ell s_2^\ell\Big)\nonumber\\
  &=&N_cq^2\left((1-s_1^\ell s_2^\ell)
  \left(1+0\times\frac{\alpha_s}\pi\right)
  +\left[\frac43\times\frac{\alpha_s}\pi s_1^\ell s_2^\ell\right]\right).  
\end{eqnarray}
In Tables~\ref{tab1} and~\ref{tab2} we list all $m_q=0$ and $m_q\to 0$
contributions to the various spin configurations for the parity-even ($VV$)
and parity-odd ($VA$) current contributions using the representation
Eq.~(\ref{anomalies}). We have again used the square bracket notation to
denote the anomalous contributions. The non-vanishing anomalous spin-flip
contribution proportional to $[4/3]$ in Table~\ref{tab1} e.g. in the 
$(\uparrow\uparrow)$ spin configuration agrees with the corresponding result 
in Ref.~\cite{Smi} where this contribution was referred to as the chirality 
breaking contribution.
\begin{table}\begin{center}
\begin{tabular}{|c||c|c|c|}\hline $VV$&$U$&$L$&$F$\\\hline\hline
$(\uparrow\uparrow)$&$0+\frac{\alpha_s}\pi(0+[4/3])$&
$0+\frac{\alpha_s}\pi(0+[0])$&
$0+\frac{\alpha_s}\pi(0+[0])$\\\hline
$(\uparrow\downarrow)$&$2+\frac{\alpha_s}\pi(2/3-[4/3])$&
$0+\frac{\alpha_s}\pi(4/3+[0])$&
$2+\frac{\alpha_s}\pi(0 -\,[4/3])$\\\hline
$(\downarrow\uparrow)$&$2+\frac{\alpha_s}\pi(2/3-[4/3])$&
$0+\frac{\alpha_s}\pi(4/3+[0])$&
$-2+\frac{\alpha_s}\pi(0+[4/3])$\\\hline
$(\downarrow\downarrow)$&$0+\frac{\alpha_s}\pi(0+[4/3])$&
$0+\frac{\alpha_s}\pi(0+[0])$&
$0+\frac{\alpha_s}\pi(0+[0])$\\\hline
\end{tabular}
\caption{\label{tab1}Born-term and $O(\alpha_s)$ corrections to specific
spin configurations (first column) in QCD($m_q=0$) and QCD($m_q\to 0$) for the
parity-even current contributions. The entries are given in terms of
contributions to the hadron tensor components
$H^{VV}_\alpha(s_1^\ell,s_2^\ell)=H^{AA}_U(s_1^\ell,s_2^\ell)$
($\alpha=U,L,F$) in units of $N_cq^2$. Anomalous contributions are shown in
square brackets.}
\end{center}\end{table}

\begin{table}\begin{center}
\begin{tabular}{|c||c|c|c|}\hline $VA$&$U$&$L$&$F$\\\hline\hline
$(\uparrow\uparrow)$&$0+\frac{\alpha_s}\pi(0+[0])$&
$0+\frac{\alpha_s}\pi(0+[0])$&
$0+\frac{\alpha_s}\pi(0+[4/3])$\\\hline
$(\uparrow\downarrow)$&$2+\frac{\alpha_s}\pi(2/3-[4/3])$&
$0+\frac{\alpha_s}\pi(4/3+[0])$&
$2+\frac{\alpha_s}\pi(0-[4/3])$\\\hline
$(\downarrow\uparrow)$&$-2+\frac{\alpha_s}\pi(-2/3+[4/3])$&
$0+\frac{\alpha_s}\pi(-4/3+[0])$&
$2+\frac{\alpha_s}\pi(0-[4/3])$\\\hline
$(\downarrow\downarrow)$&$0+\frac{\alpha_s}\pi(0+[0])$&
$0+\frac{\alpha_s}\pi(0+[0])$&
$0+\frac{\alpha_s}\pi(0+[4/3])$\\\hline
\end{tabular}
\caption{\label{tab2}Born-term and $O(\alpha_s)$ corrections to specific spin
configurations (first column) in QCD($m_q=0$) and QCD($m_q\to 0$) for the
parity-odd current contributions. The entries are given in terms of
contributions to the hadron tensor components
$H^{VA}_U(s_1^\ell,s_2^\ell)=H^{AV}_U(s_1^\ell,s_2^\ell)$,
$H^{VA}_L(s_1^\ell,s_2^\ell)=H^{AV}_L(s_1^\ell,s_2^\ell)$, and
$H^{VA}_F(s_1^\ell,s_2^\ell)=H^{AV}_F(s_1^\ell,s_2^\ell)$ in units of
$N_cq^2$. Anomalous contributions are shown in square brackets.}
\end{center}\end{table}

Using the $m_q\to 0$ limiting expressions Eqs.~(\ref{helimit1})
and~(\ref{helimit2}), one finally obtains for the longitudinal spin--spin
asymmetry
\begin{eqnarray}\label{anocorr}
P^{\ell\ell}(\cos\theta)&=&-1+[4]\frac{\alpha_sC_F}{4\pi}\left(1
  +\left(\frac{(1+\cos^2\theta+4\sin^2\theta)g_{11}}{(1+\cos^2\theta)g_{11}
  +2\cos\theta g_{44}}\right)\frac{\alpha_sC_F}{4\pi}\right)^{-1} \nonumber\\
  &=&-1+[4]\frac{\alpha_sC_F}{4\pi}+O(\alpha_s^2).
\end{eqnarray}
where we again have encased the ``$4$'' in square brackets in order to
identify the contribution as anomalous. The result in Eq.~(\ref{anocorr}) is
independent of the flavour of the quark and thereby also holds for polarized
lepton pair production $e^+e^-\to\mu^+\mu^-,\tau^+\tau^-$ replacing, of course,
$C_F\alpha_s$ by $\alpha$. We emphasize again that the normal $m_q=0$
contributions lead to the value $P^{\ell\ell}(\cos\theta)=-1$ in the high
energy limit. It is quite remarkable that the anomalous contributions do not
show any $\cos\theta$ dependence at order $O(\alpha_s)$.


\section{Anomalous contributions from the universal splitting function}


It was shown e.g.\ in Ref.~\cite{Falk} that, in the limit $m_e\to 0$, there is
a finite helicity flip radiation (i.e.\ $e_L\to e_R+\gamma$ and
$e_R\to e_L+\gamma$) which arises from collinear photon emission. It can be
described in terms of a universal splitting function
$D_{hf}^{\rm anom}=\frac\alpha{2\pi}z$ where $z$ is the fractional energy of
the photon (not to be confused with the energy-type variable used in this
paper). In this order of perturbation theory the derivation of
Ref.~\cite{Falk} can be directly transcribed to the QCD case where one now has
\begin{equation}\label{anomsplit}
D_{hf}^{\rm anom}=C_F\frac{\alpha_s}{2\pi}z.
\end{equation}
Integrating over the fractional energy $z$ gives
\begin{equation}
H_{hf}^{\rm anom}=\int_0^1H({\it Born})C_F\frac{\alpha_s}{2\pi}z\,dz=
C_F\frac{\alpha_s}{4\pi}H({\it Born})
\end{equation}
where we have, for the sake of simplicity, dropped all indices on the
hadron tensor $H({\it Born\/})$.

One knows that there is no anomalous contribution to a given fermion transition
when one integrates over the whole two-body phase space and one sums over the
two spin states of a given final fermion. This was explicitly demonstrated for
muon decay
$\mu^-\to e^-\,\bar\nu_e\,\nu_\mu$ in Ref.~\cite{Fischer:2002hn} where the
$O(\alpha)$ contributions to the two final electron spin states
$e_L\to e_R+\gamma$ and $e_L\to e_L+\gamma$ were calculated in the limit
$m_e\to 0$. There was an anomalous spin-flip contribution to
$e_L\to e_R+\gamma$ determined by the universal splitting function in
Eq.~(\ref{anomsplit}) accompanied by an anomalous non-flip contribution to
$e_L\to e_L+\gamma$ of equal and opposite strength. When summing the
spin-flip and non-flip contributions the anomalous terms cancelled leading to
a ``normal'' unpolarized rate. This implies that, when integrating over the
whole phase space, every anomalous spin-flip contribution is accompanied by an
anomalous non-flip contribution of equal and opposite strength. As shown e.g.\
in Ref.~\cite{Falk} the photon in the collinear emission process
$e_L\to e_R+\gamma$ is lefthanded, i.e.\ one has $e_L\to e_R+\gamma_L$ in
agreement with angular momentum conservation. By the same token the anomalous
non-flip contribution must vanish in the forward direction as has been
explicitly demonstrated in Ref.~\cite{Falk}.

The same empirical pattern is observed in the $m_q\to 0$ limit of
$e^+e^-\to q\bar qg$. Consider, for example, gluon emission from the quark and
keep the antiquark helicity fixed at $\bar q_L$. The anomalous spin-flip
contribution $q_L\to q_R+g$ is determined by the universal splitting function
Eq.~(\ref{anomsplit}) which is cancelled by the residual non-flip contribution
$q_L\to q_L+g$ when one takes the sum of the two, i.e.\ the sum has no
residual mass effects. We have verified by explicit calculation that the
residual non-flip contribution originates from the near-forward region
irrespective of the fact that the non-flip contribution vanishes in the exact
forward direction.

In this way one can determine the anomalous flip and non-flip contributions to
each spin configuration. In fact, one reproduces the anomalous entries in
Table~\ref{tab1} and Table~\ref{tab2} by using the relations
\begin{eqnarray}
\qquad[\uparrow\uparrow]&=&(\uparrow\downarrow_{[hf]})
  +(\downarrow_{[hf]}\uparrow)\nonumber\\
  &=&\Big((\uparrow \downarrow)_{\it Born}
  +(\downarrow \uparrow)_{\it Born}\Big)\left[C_F\frac{\alpha_s}{4\pi}\right]
\end{eqnarray}
and
\begin{eqnarray}
\hspace{-2.0cm}[\uparrow\downarrow]&=&(\uparrow\downarrow_{[nf]})
  +(\uparrow_{[nf]}\downarrow) \nonumber \\
  &=&-2(\uparrow\downarrow)_{\it Born}\left[C_F\frac{\alpha_s}{4\pi}\right],
\end{eqnarray}
and correspondingly for the spin configurations $[\downarrow\uparrow]$
and $[\downarrow\downarrow]$. The anomalous contributions to the spin--spin
structure function $H_U^{\ell_1\ell_2}$ can be seen to be $50\%$ flip and
$50\%$ non-flip. The anomalous contribution to the single-spin structure
function $H_U^{\ell_1}$ derives entirely from the anomalous non-flip
contribution contrary to what was stated in~\cite{GKT2}. This, in particular,
means that the universal anomalous spin-flip contribution will not show up in
single-spin polarization measurements, but contributes at the $50\%$ level to
spin--spin polarization measurements.

One can explicitly verify from the entries of Table~\ref{tab1} and
Table~\ref{tab2} that the anomalous flip and non-flip contributions to a given
quark or antiquark transition cancel upon summation. We mention that the
``normal'' $O(\alpha_s)$\, $m_q=0$ contributions in Table~\ref{tab1} and
Table~\ref{tab2} can be determined from Eq.~(\ref{anomalies}).


\section{Numerical results}


We are now in the position to discuss our numerical results for the
$\cos\theta$-dependent normalized longitudinal spin--spin asymmetry
$P^{\ell\ell}(\cos\theta)$ defined in Eq.~(\ref{corrdef}).

\begin{figure}
\epsfig{figure=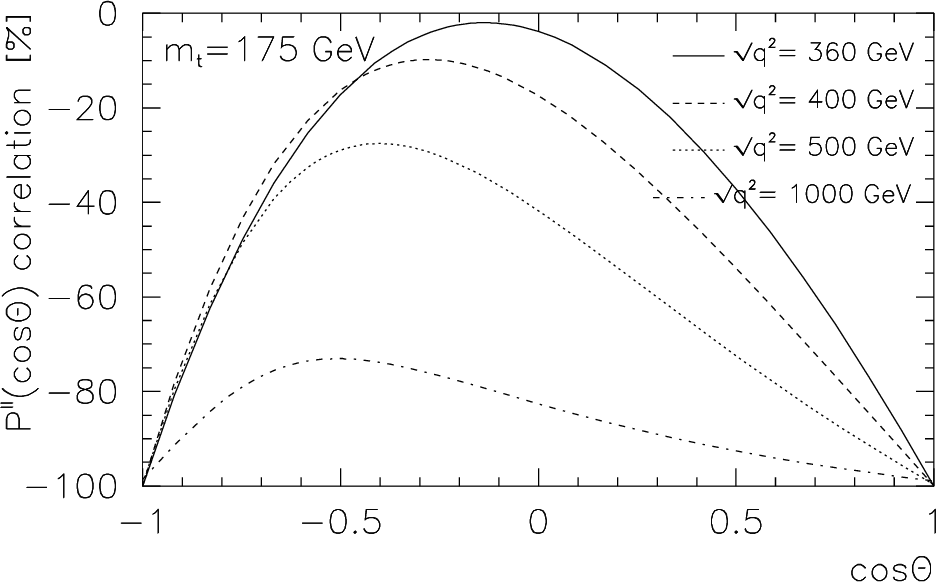, scale=0.8}
\caption{\label{fig1}$O(\alpha_s)$ polar angle dependence of the longitudinal
  spin--spin correlation asymmetry in $e^+e^-\to t\bar t$ at $\sqrt{q^2}=360$,
$400$, $500$, and $1000\GeV$. We always take $m_t=175\GeV$ as default value.}
\end{figure}

In Fig.~\ref{fig1} we show the $\cos\theta$ dependence of the $O(\alpha_s)$
longitudinal spin--spin correlation function $P^{\ell\ell}(\cos\theta)$ in
$e^{+}e^{-}\to t\bar t$ for the four different center-of-mass energies
$\sqrt{q^2}=360$, $400$, $500$, and $1000\GeV$. At $\sqrt{q^2}=360\GeV$ one is
sufficiently far above the $(t\bar t)$ threshold for perturbative QCD to apply.
The above range of center-of-mass energies is the envisaged range for the
proposed linear colliders. At the lowest shown energy $\sqrt{q^2}=360\GeV$ the
$\cos\theta$ distribution of $P^{\ell\ell}$ is still quite close to the
threshold Born-term distribution $P^{\ell\ell}(\cos\theta)=-\cos^2\theta$
(see Eq.~(\ref{pllborn2})). However, one already notes a small
forward-backward asymmetric effect which is induced by the $VA$ interference
term $H_F^{4\,\{\ell_1\ell_2\}}$. In the forward and backward directions
$\cos\theta=\pm1$ the longitudinal spins are quite close to being $100\%$
aligned as is true for the Born-term case (see Eq.~(\ref{pllborn1})). Away
from the forward and backward direction the destructive longitudinal
contribution $H_L^{j\,(\ell_1\ell_2)}$ ($j=1,2$) comes into play and reduces
$P^{\ell\ell}(\cos\theta)$ from the maximal $m_q=0$ value
$P^{\ell\ell}\simeq-1$. As the energy is increased, $H_L^{j\,(\ell_1\ell_2)}$
becomes smaller and thus $P^{\ell\ell}(\cos\theta)$ becomes flatter. At
$\sqrt{q^2}=1000\GeV$ the alignment of the spins of the top and antitop
exceeds $\simeq75\%$ over most of the range of $\cos\theta$.

\begin{figure}
\epsfig{figure=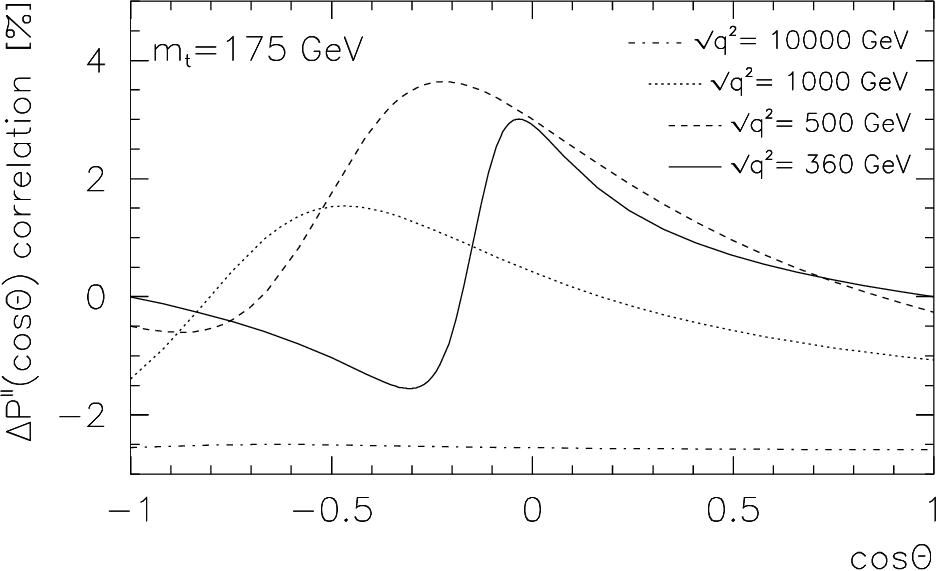, scale=0.8}
\caption{\label{fig2}Relative size of $O(\alpha_s)$ corrections to the
  longitudinal spin--spin correlation asymmetry in $e^+e^-\to t\bar t$ at
$\sqrt{q^2}=360$, $500$, $1000$, and $10000\GeV$}.
\end{figure}
\begin{figure}
\epsfig{figure=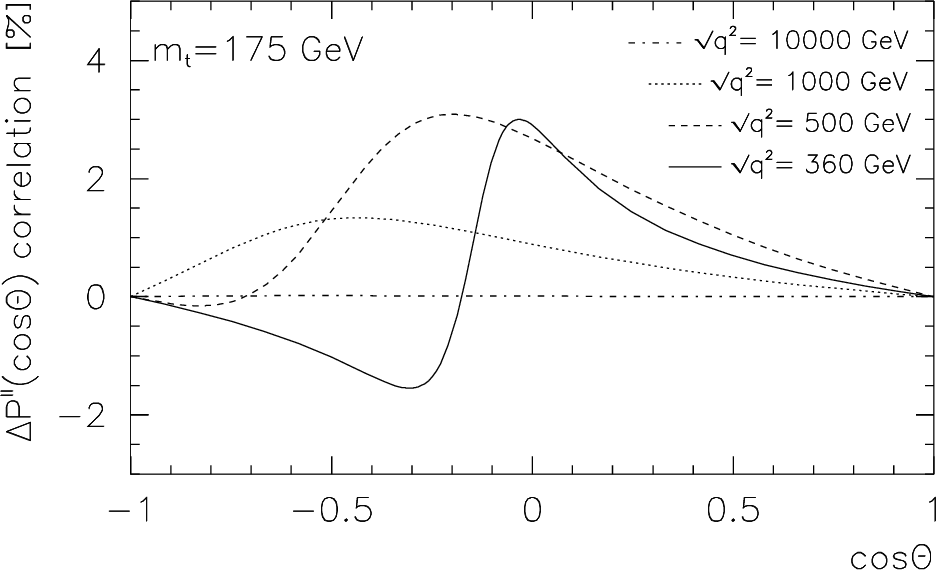, scale=0.8}
\caption{\label{fig3}Relative size of $O(\alpha_s)$ corrections to the
  longitudinal spin--spin correlation asymmetry in $e^+e^-\to t\bar t$ in
the soft gluon approximation at
$\sqrt{q^2}=360$, $500$, $1000$, and $10000\GeV$}.
\end{figure}

In Fig.~\ref{fig2} we exhibit the relative size of the $O(\alpha_s)$
corrections which we define by the measure
\begin{equation}\label{relcorr}
\Delta P^{\ell\ell}(\cos\theta)=\frac{P^{\ell\ell}(\cos\theta)|_{O(\alpha_s)}
  -P^{\ell\ell}(\cos\theta)|_{\it Born}}{P^{\ell\ell}(\cos\theta)|_{\it Born}}.
\end{equation}
Fig.~\ref{fig2} shows that the $O(\alpha_s)$ corrections to
$P^{\ell\ell}(\cos\theta)$ are strongly polar angle dependent. They are
not very small and can reach values as high as $\simeq 4\%$. It is quite
remarkable that the radiative corrections for the lowest energy
$\sqrt{q^2}=360\GeV$ are very close to zero in the forward and backward
directions. As the energy is increasing the radiative corrections in the
forward and backward directions become larger and they start approaching their
anomalous limiting value Eq.~(\ref{anocorr}).

In Fig.~\ref{fig2} we have also included a plot of
$\Delta P^{\ell\ell}(\cos\theta)$ for the rather large center-of-mass energy
of $\sqrt{q^2}=10^4\GeV$ (we are not implying that such center-of-mass
energies will ever be reached in terrestrial laboratories). The purpose is to
exhibit the anomalous contributions induced by the $O(\alpha_s)$ corrections
when $m_q/\sqrt{q^2}\to 0$. In terms of the measure
$\Delta P^{\ell\ell}(\cos\theta)$ defined in Eq.~(\ref{relcorr}) one has in
the high energy limit (see Eq.~(\ref{anocorr}))
\begin{equation}
\Delta P^{\ell\ell}(\cos\theta)=-[4]\frac{\alpha_s}{3\pi}.
\end{equation}
Using $\alpha_s(\sqrt{q^2}=10^4\GeV)=0.0712$ one numerically has
$\Delta P^{\ell\ell}(\cos\theta)=-0.0303$ which is quite close to the value
$\simeq-0.026$ shown in Fig.~\ref{fig2}. The remaining difference can be
traced to non-asymptotic effects.

In order to highlight the importance of hard gluon emission for the
radiatively corrected longitudinal spin--spin asymmetry we have determined
$\Delta P^{\ell\ell}(\cos\theta)$ in the soft gluon approximation as has also
been employed in Ref.~\cite{Akatsu:1997tq}. Fig.~\ref{fig3} shows a
plot of $\Delta P^{\ell\ell}(\cos\theta)$ in the soft gluon approximation
with the gluon energy integrated to its maximal energy. Since the soft
gluon approximation is essentially a two-body approximation the radiative
corrections can be seen to vanish in the forward and backward directions as
discussed in Sec.~3. For the lowest energy $\sqrt{q^2}=360\GeV$ the
corresponding two curves in Figs.~\ref{fig2} and~\ref{fig3} practically
lie on top of each other since there is very little phase space left for
gluon emission. At $\sqrt{q^2}=500\GeV$ the difference between the full and
approximate calculations becomes noticeable. In particular the
$\cos\theta$ dependence of the full calculation is more pronounced. The same
holds true for $\sqrt{q^2}=1000\GeV$ where the difference between the full and
approximate calculation becomes as large as $\approx 1\%$ in the forward and
backward directions. At $\sqrt{q^2}=10000\GeV$ one has
$\Delta P^{\ell\ell}(\cos\theta)\approx 0$ in the soft gluon approximation as
expected whereas $\Delta P^{\ell\ell}(\cos\theta)\approx -2.6\%$ in the full
calculation due to the anomalous contributions as discussed before. 

\begin{figure}
\epsfig{figure=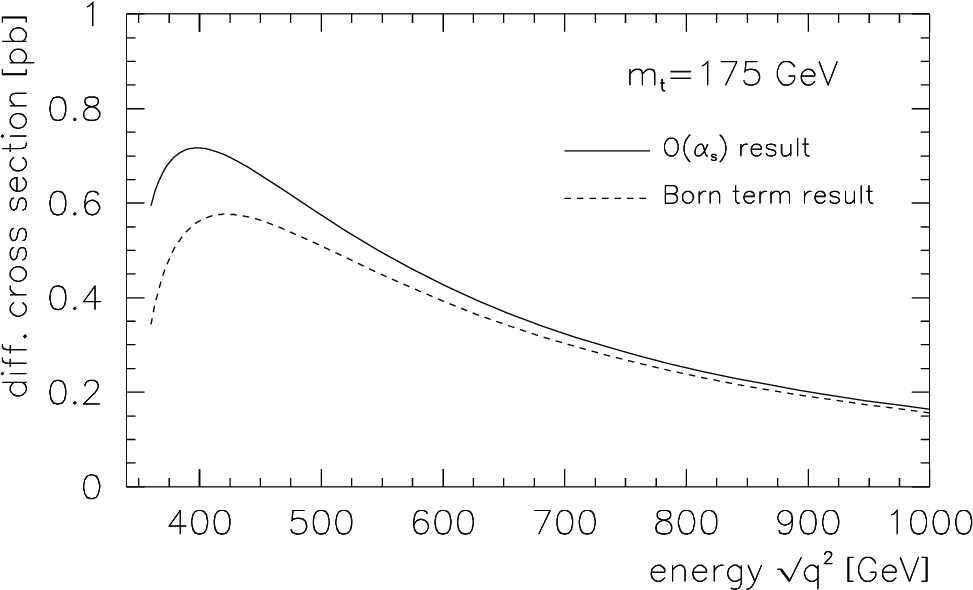, scale=0.8}
\caption{\label{fig4}Differential cross section for $e^+e^-\to t\bar t$ at
  different center-of-mass energies $\sqrt{q^2}$. Shown are Born-term results
  (dashed line) and $O(\alpha_s)$ results (solid line)}.
\end{figure}

The relatively small value of the $O(\alpha_s)$ correction for the
longitudinal spin--spin asymmetry in Fig.~\ref{fig2} has to be contrasted with
the large $O(\alpha_s)$ corrections to the cross section as shown in
Fig.~\ref{fig4}. For example, at $\sqrt{q^2}=500\GeV$ the $O(\alpha_s)$
corrections to the cross section amount to $15.4\%$. The reason that the
radiative corrections to $P^{\ell\ell}$ are smaller than those for the rate is
that the $O(\alpha_s)$ corrections in the numerator and denominator of
$P^{\ell\ell}$ tend to go in the same direction and thus tend to cancel when
one takes the ratio. However, this is not true for the anomalous contribution
which contributes only to the numerator. 

It is quite remarkable that the radiative corrections to $P^{\ell\ell}$ in the
forward and backward directions are not small for the higher energies even if
they only arise from non-Born-term-like hard gluon emission. Compare this to
the corresponding case of top quark decay $t\to b+W_R^+$ where the
right-handed $W_R^+$ is again only populated by non-Born-term-like hard gluon
emission. In this case the radiative corrections amount to only $0.1\%$
of the total Born-term rate~\cite{Fischer:1998gsa}. The explanation is that in
the $e^{+}e^{-}\to t\bar t$ case one is sensitive to the enhanced forward 
hard gluon emission region whereas in the $t\to b+W_R^+$ case there is no such
enhancement.

\begin{figure}
\epsfig{figure=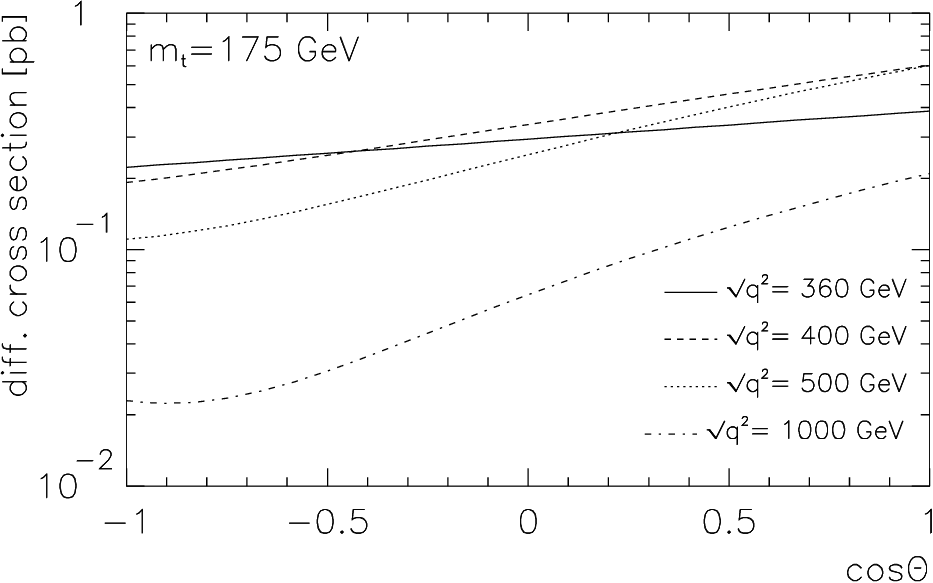, scale=0.8}
\caption{\label{fig5}$O(\alpha_s)$ polar angle distribution of the production
  rate in $e^+e^-\to t\bar t$ at $\sqrt{q^2}=360$, $400$, $500$, and
  $1000\GeV$}
\end{figure}

In Fig.~\ref{fig5} we display the polar angle dependence of the $O(\alpha_s)$
differential cross section for $(t\bar t)$-production in $e^+e^-$-annihilation.
As Fig.~\ref{fig5} shows, the cross section strongly peaks in the forward
direction. For example, for $\sqrt{q^2}=1000\GeV$ there is more than a factor
of $10$ cross section difference between the forward and backward directions.
The fact that the cross sections peak so strongly in the forward direction is
quite welcome from the point of view of checking the $O(\alpha_s)$ prediction
for the longitudinal spin--spin asymmetry which, as remarked on before, is
close to its $m_q=0$ value $P^{\ell\ell}=-1+4\alpha_s/3\pi$ in the forward
direction.

\begin{figure}
\epsfig{figure=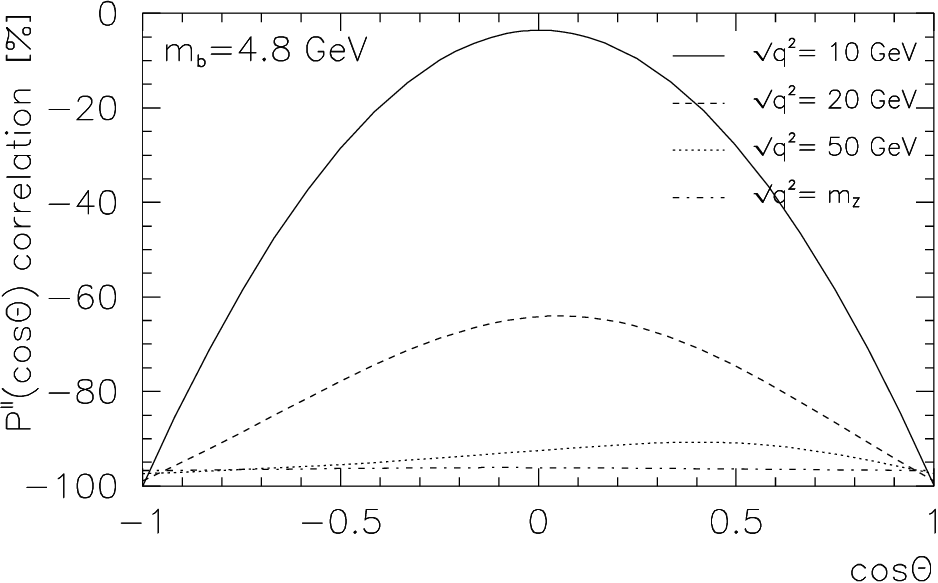, scale=0.8}
\caption{\label{fig6}Polar angle dependence of the longitudinal spin--spin
  asymmetry in $e^+e^-\to b\bar b$ at $\sqrt{q^2}=10$, $20$ and $50\GeV$ and
  on the $Z$ peak}
\end{figure}

For the sake of completeness we also give corresponding results for the
longitudinal spin--spin asymmetry in the bottom quark sector. The bottom quark
is much longer lived than the top quark and has ample time to hadronize before
it decays. Spin information is completely lost when the bottom quark fragments
into bottom mesons~\cite{Close:1992fw,Falk:1993rf}. However, for fragmentation
into bottom baryons, which is expected to occur at the $10\%$ level, one can 
anticipate a polarization transfer of $70\%$ from the bottom quark to the
final $\Lambda_b$ baryon~\cite{Falk:1993rf}. The problem of how to best
measure the $\Lambda_b$ polarization has been addressed in
Refs.~\cite{Bonvicini:1994mr,Diaconu:1995mp}. 

Concerning rates at the ILC let us focus on a c.m.\ energy of $500\GeV$ where 
the $(b\bar b)$-rate approximately equals the ($t\bar t)$-rate of $\sim0.5\,pb$
(see e.g.\ Ref.~\cite{GKT1}). For a projected ILC luminosity of
$L=2\cdot10^{34}cm^{-2}s^{-1}$ one then would obtain $3\cdot10^3$
$(\Lambda_b\bar\Lambda_b)$-pairs in a year, which optimistically would allow
one to perform the proposed spin--spin correlation measurements. Much higher
$(\Lambda_b\bar\Lambda_b)$ production rates can be expected at the GigaZ
option (see e.g.\ Ref.~\cite{Erler:2000jg}) if it should be realized at the
ILC. One expects a yield of $10^9$ $Z$-bosons per year at the GigaZ which will
decay to $0.14\times10^9$ $(b\bar b)$ pairs resulting in a final yield of
$\sim 1.4\cdot 10^6$ $(\Lambda_b\bar\Lambda_b)$-pairs in a year. 

\begin{figure}
\epsfig{figure=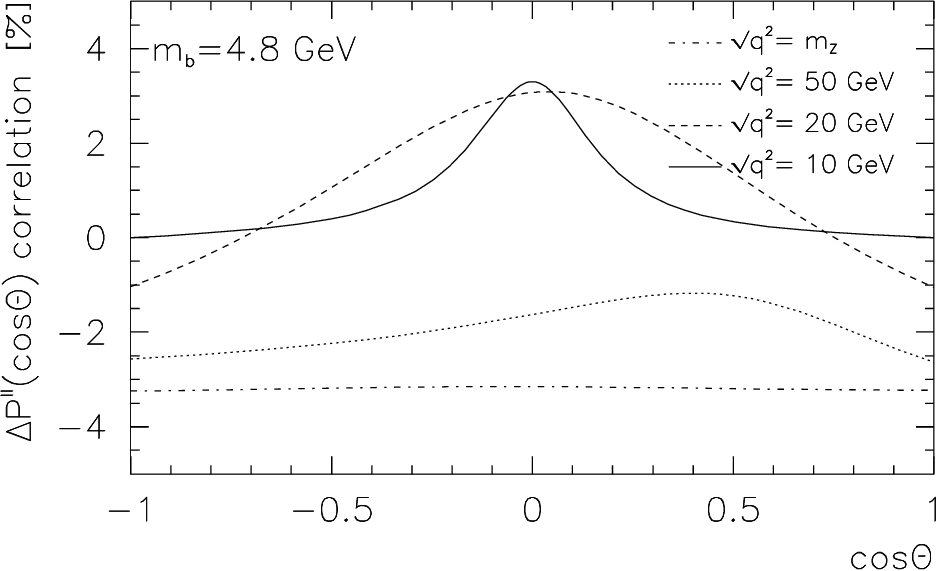, scale=0.8}
\caption{\label{fig7}Relative size of $O(\alpha_s)$ corrections to the
  longitudinal spin--spin asymmetry in $e^+e^-\to b\bar b$ at $\sqrt{q^2}=10$,
  $20$ and $50\GeV$ and on the $Z$ peak}
\end{figure}

\begin{figure}
\epsfig{figure=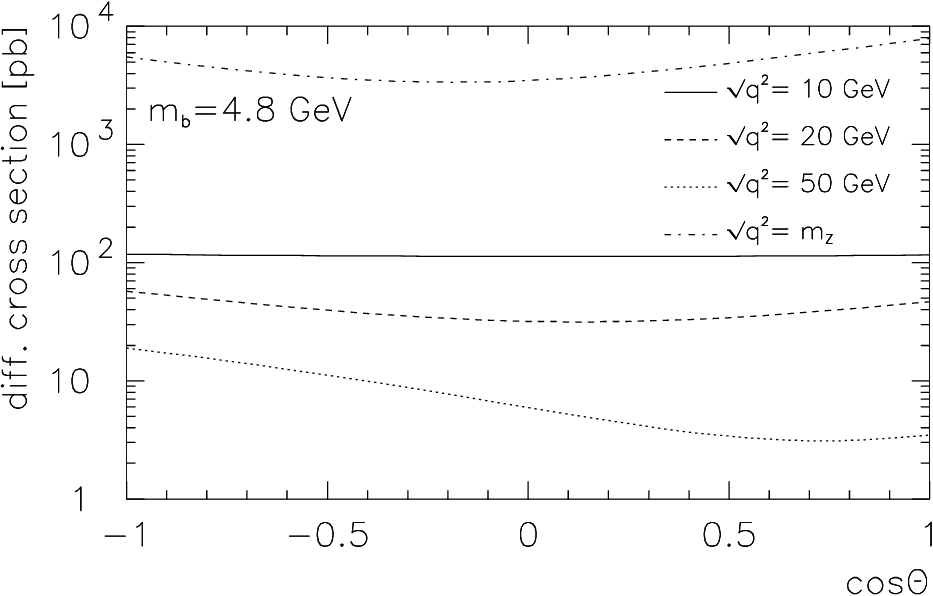, scale=0.8}
\caption{\label{fig8}$O(\alpha_s)$ polar angle distribution of the production
  rate in $e^+e^-\to b\bar b$ at $\sqrt{q^2}=10$, $20$ and $50\GeV$ and on the
  $Z$ peak}
\end{figure}

In Figs.~\ref{fig6}--\ref{fig8} we show plots for $(b\bar b)$-production at
center-of-mass energies of $\sqrt{q^2}=10$, $20$ and $50\GeV$ and at the $Z$
pole ($m_Z=91.188\GeV$). We use a pole mass of $m_b=4.8\GeV$ for the bottom
quark. In Fig.~\ref{fig6} we show the polar angle dependence of the
$O(\alpha_s)$ longitudinal spin--spin correlation function
$P^{\ell\ell}(\cos\theta)$. The forward-backward asymmetry of $P^{\ell\ell}$
is not as pronounced as in the top quark case and, contrary to the top quark
case, its angular distribution is only slightly enhanced in the forward
hemisphere. At the $Z$ pole one has an almost flat distribution close to the
asymptotic value of $P^{\ell\ell}=-1+4\alpha_{s}/3\pi$. The deviation from
the $-100\%$ value is again due to the anomalous contribution.

In Fig.~\ref{fig7} we exhibit the size of the $O(\alpha_s)$ corrections
defined by the measure $\Delta P^{\ell\ell}$ as defined in Eq.~(\ref{relcorr}).
Close to threshold at $\sqrt{q^2}=10\GeV$ the $O(\alpha_s)$ corrections in the
forward and the backward directions are quite small. The corrections become
larger away from the forward and backward directions and rise to $\simeq3.5\%$
at $\theta=90^0$. As the energy increases, the $O(\alpha_s)$ corrections become
progressively flatter and tend to turn to negative values. At the $Z$ peak the
$O(\alpha_s)$ corrections are almost flat and reach a value of $\simeq-3.4\%$
which again shows that they are dominated by the anomalous contribution.

In Fig.~\ref{fig8} we show the $\cos\theta$ dependence of the
$(b\bar b)$-production cross section. The distributions are much flatter than
in the top quark pair production case. The differential cross section falls as
the center-of-mass energy increases and then reaches back to high values on
the $Z$ peak.


\section{Summary and conclusion}


We have presented analytical results for the $O(\alpha_s)$ corrections to the
polar angle-dependent longitudinal spin--spin asymmetry
$P^{\ell\ell}(\cos\theta)$ in $e^+e^-\to q\bar q$. A numerical evaluation of
$P^{\ell\ell}(\cos\theta)$ for top quark pair production shows that the
$O(\alpha_s)$ corrections are not small and, depending on the energy, can be
strongly polar angle dependent. The $O(\alpha_s)$ corrections to
$P^{\ell\ell}$ are smaller than the $O(\alpha_s)$ corrections to the
cross section itself. The reason is that the $O(\alpha_s)$ corrections to the
numerator and the denominator in the (normalized) longitudinal spin--spin
asymmetry tend to go in the same direction with the effect that the
$O(\alpha_s)$ correction to the ratio becomes reduced. At lower energies the
longitudinal spins of the quark and antiquark become
almost $100\%$ aligned in the forward and backward direction, i.e.\ one has
$P^{\ell\ell}(\cos\theta=\pm1)\simeq-1$ with small radiative corrections. As
the energy is increased the radiative corrections become larger at the
endpoints signalling a slow approach to the asymptotic regime.

For bottom quark pair production on the $Z$ peak one is already close to the
asymptotic regime since $m_b^2/m_Z^2=2.8\times 10^{-3}$ (using $m_b=4.8\GeV$,
$m_Z=91.188\GeV$ as before). In massless QCD one expects
$P^{\ell\ell}(\cos\theta)=-1$ to any order of $\alpha_s$ because there are no
spin-flip contributions in massless QCD. This is different in the $m_q\to 0$
(or high energy) limit of QCD where there are residual mass effects starting
at $O(\alpha_s)$ which change the naive pattern of massless QCD. In fact,
in the high energy limit (or for $m_q\to 0$) one obtains instead
$P^{\ell\ell}(\cos\theta)=-1+4\alpha_s/3\pi$ due to the anomalous
$O(\alpha_s)$ flip and non-flip contributions. The numerical evaluation of the
longitudinal spin--spin asymmetry for bottom quark pair production on the $Z$
peak shows that one is already close to the asymptotic form but noticeable
preasymptotic effects are still present with respect to the flatness of the
$\cos\theta$ distribution and the size of the anomalous contributions. We have
discussed in some detail the role of the anomalous contributions and how they
contribute to spin-flip and non-flip terms in the three correlation structure
functions that determine the $\cos\theta$ dependence of the longitudinal
spin--spin asymmetry.

All our results can be applied to QED final state corrections in polarized
lepton pair production in $e^+e^-\to\mu^+\mu^-$ and $e^+e^-\to\tau^+\tau^-$.
The radiative corrections are, of course, smaller in the QED case due to the
smallness of the QED coupling constant $\alpha$. In particular, in the
asymptotic limit, where $P^{\ell\ell}(\cos\theta)=-1+\alpha/\pi$, the
anomalous contribution to the longitudinal spin--spin asymmetry is now only of
$O(0.23\%)$.

What remains to be done is to calculate the radiative corrections to the
remaining transverse-longitudinal and transverse-transverse spin--spin
density matrix elements in the helicity system. Alternatively, one can study
the spin--spin density matrix in the off-diagonal basis representation of
Ref.~\cite{ps96}, which is optimal at the Born-term level. Radiative
corrections to the spin--spin density matrix in the off-diagonal basis have
been considered before using the soft gluon approximation~\cite{Akatsu:1997tq}.
It is, however, clear that, in the soft gluon approximation, one misses
potentially large contributions from e.g.\ the near-forward emission of hard
gluons as described in this paper.

\vspace{12pt}\noindent
{\bf Acknowledgements:} The work of S.~G. is supported by the Estonian
target financed project No.~0182647s04, by the Estonian Science Foundation
under grant No.~6216 and by the Deutsche Forschungsgemeinschaft (DFG)
under grant 436 EST 17/1/06.

\end{document}